\documentclass[aps,prd,twocolumn,nofootinbib,float]{revtex4-1}

\usepackage{booktabs,makecell,multirow}
\usepackage{amsmath}
\usepackage{ulem}
\usepackage{booktabs}

\usepackage{graphicx,epsfig}

\usepackage[colorlinks=true, linkcolor=blue, citecolor=blue, urlcolor=blue]{hyperref}

\begin{document}

\title{Determination of $\alpha_s(M_Z)$ via a high-precision effective coupling $\alpha^{g_1}_s(Q)$}

\author{Qing Yu$^{1}$}
\email{yuq@swust.edu.cn}

\author{Xing-Gang Wu$^{2}$}
\email{wuxg@cqu.edu.cn}

\author{Hua Zhou$^{1}$}
\email{zhouhua@swust.edu.cn}

\author{Jian-Ming Shen$^{3}$}
\email{shenjm@hnu.edu.cn}

\affiliation{$^1$School of Mathematics and Physics, Southwest University of Science and Technology, Mianyang 621010, P.R. China}

\affiliation{$^2$Department of Physics, Chongqing Key Laboratory for Strongly Coupled Physics, Chongqing University, Chongqing 401331, P.R. China}

\address{$^3$ School of Physics and Electronics, Hunan University, Changsha 410082, P.R. China}

\date{\today}

\begin{abstract}

We propose a novel method to determine the strong coupling of quantum chromodynamics (QCD) and fix its running behavior at all scales by using the Bjorken sum rules (BSR). The BSR defines an effective coupling $\alpha^{g_1}_s(Q)$ which includes the nonperturbative high-twist corrections and perturbative QCD (pQCD) corrections to the leading-twist part. For the leading-twist part of $\alpha^{g_1}_s(Q)$, we adopt the infinite-order scale-setting procedure of the principle of maximum conformality ($\rm{PMC}_\infty$) to deal with its pQCD corrections, which reveals the intrinsic conformality of series and eliminates conventional renormalization scheme-and-scale ambiguities. Using the $\rm{PMC}_\infty$ approach, we not only eliminate \textit{the first kind of residual scale dependence} due to uncalculated higher-order terms, but also resolve the previous ``self-consistence problem". The holographic light-front QCD model is used for $\alpha^{g_1}_s(Q)$ in the infrared region, which also reveals a conformal behavior at $Q\to 0$. As a combination, we obtain a precise $\alpha^{g_1}_s(Q)$ at all scales, which matches well with the known experimental data with $p$-value $\sim99\%$, we determine the strong coupling constant at the critical scale $M_Z$, $\alpha_s(M_Z)=0.1191\pm{0.0012}\mp0.0006$, where the first error comes from $\Delta\kappa$ of LFHQCD model and the second error is from \textit{the second kind of residual scale dependence} that is negligible.

\end{abstract}

\maketitle

\section{Introduction}

The Quantum Chromodynamics (QCD) strong coupling constant $\alpha_s$ is one of the fundamental parameters in QCD and it indicates the strength of gluon-quark interactions. Knowing its precise magnitude is crucial for attaining highly precise QCD predictions and exploring the intricate internal structure of hadrons. The strong coupling constant $\alpha_s$ is scale dependent, and due to the asymptotic freedom inherent in QCD theory~\cite{Gross:1973id, Politzer:1973fx}, its magnitude progressively diminishes as the scale increases and exhibits a small value when the scale is much larger than the critical QCD scale $\Lambda_{\rm QCD}$. On this basis, the perturbative QCD (pQCD) corrections to physical observables can be routinely expanded as a perturbative series with increasing $\alpha_s$-powers. It is then important to know the magnitude $\alpha_s$ at any scales. Practically, one usually extracts the magnitude of $\alpha_s$ at some typical energy scale such as the $Z$-boson mass $M_Z$ via comparison with experimental data, and then determines its value at any scale by using the following renormalization group equation (RGE)~\cite{tHooft:1973mfk, Weinberg:1973xwm},
\begin{eqnarray}
\mu^2 {\partial \over\partial\mu^2}\left({\alpha_s(\mu)\over\pi}\right) =\beta(\alpha_s)=-{\beta_i}\left({\alpha_s(\mu)\over\pi}\right)^{i+2},
\label{RGE}
\end{eqnarray}
where $\mu$ stands for the energy scale, $\alpha_s$ and the perturbatively calculable $\{\beta_i\}$-functions are generally scheme-dependent, whose first several terms under the $\overline{\rm MS}$-scheme and $V$-scheme are put in Appendix. And to achieve a reliable and precise pQCD prediction, it is also imperative to find an appropriate method for determining the correct effective $\alpha_s$ (and hence the corresponding effective scale) of the considered process.

The Bjorken Sum rule (BSR) $\Gamma^{p-n}_1(Q)$~\cite{Bjorken:1966jh, Bjorken:1969mm} describes the differences between the spin structure function of the proton and neutron, $g^{p,n}_1(x)$, with the Bjorken scaling variable $x$ and the energy scale $Q$. The measurable physical observable $\Gamma^{p-n}_1(Q)$ provides a platform for extracting the strong coupling constant $\alpha_s$ at various energy scales. Experimentally, the HERMES Collaboration at DESY~\cite{Ackerstaff:1997ws}, the COMPASS Collaboration at CREN~\cite{Alexakhin:2006oza}, the E142 and E143 Collaborations at SLAC~\cite{Anthony:1993uf} and Deur etal. at Jefferson lab (JLab)~\cite{Deur:2004ti, Deur:2005cf, Deur:2008ej, Deur:2014vea} issued the data for $\Gamma^{p-n}_1(Q)$ within the $Q^2$-range of $0.05~{\rm GeV^2}$ to $10~{\rm GeV^2}$. Recently, Deur etal. incorporates new data within the range of $Q^2\in[0.02,4.75] \rm GeV^2$~\cite{Deur:2022msf}, thereby facilitating the extraction $\alpha_s$ with higher precision.

Theoretically, one defines an effective coupling $\alpha^{g_1}_s(Q)$ for the BSR $\Gamma^{p-n}_1(Q)$,
\begin{eqnarray}
\Gamma^{p-n}_1(Q)&=&\int^1_0 dx\left[g^p_1(x)-g^n_1(x)\right]   \nonumber\\
&=&\left|\frac{g_A}{g_V}\right|\frac{1}{6}\left[1-a^{g_1}_s(Q) \right],
\label{gammapn}
\end{eqnarray}
where $a^{g_1}_s(Q)= {\alpha^{g_1}_s(Q)} /{\pi}$, and the ratio of nucleon axial charge $\left|g_A/g_V\right|=1.2754\pm0.0013$~\cite{ParticleDataGroup:2024cfk}. The effective coupling $a^{g_1}_s(Q)$ contains not only perturbatively calculable leading-twist terms but also the non-perturbative high-twist terms. The non-perturbative high-twist terms are in the forms of $1/Q^{2n}$ under the Operator Product Expansion form~\cite{Bjorken:1966jh, Bjorken:1969mm}. On the one hand, when $Q$ is large enough, the effective coupling $a^{g_1}_s(Q)$ will be dominated by perturbative contributions and can be expanded as power series over $\alpha_s$. Till now, the pQCD corrections to $a^{g_1}_s(Q)$ in the massless limit has been calculated up to $\mathcal{O}(\alpha^4_s)$ level under the $\overline{\rm MS}$-scheme~\cite{Gorishnii:1985xm, Larin:1991tj, Baikov:2010je}, and the singlet contributions to BSR has been given up to $\mathcal{O}(\alpha^4_s)$-level in Refs.\cite{Larin:2013yba, Baikov:2015tea}. Beyond the massless limit, the heavy quark mass ($m_c$ and $m_b$) corrections to $a^{g_1}_s(Q)$ have been calculated up to $\mathcal{O}(\alpha^2_s)$-level~\cite{Blumlein:2016xcy}. On the other hand, the high-twist terms will have sizable or even dominant contributions in middle and low $Q$-range. Refs.\cite{Deur:2008ej, Deur:2014vea, Ayala:2018ulm, Yu:2021ofs, Ayala:2023cxm} extracted the high-twist corrections by comparing the JLab data with the leading-twist term of BSR.

In low-energy region, one of the successful nonperturbative approaches is the light-front holographic QCD (LFHQCD)~\cite{Brodsky:2014yha}, an approximate conformal theory describing the QCD. The LFHQCD approach is based on the AdS/CFT duality and light-front quantization. It incorporates a Gaussian $e^{+\kappa^2 z^2}$-term to modify the action of the 5-dimensional ${\rm AdS}$, where the mass scale $\kappa$ serves as a confinement scale as well. In the momentum space, the LFHQCD suggests $\alpha^{g_1,{\rm LFHQCD}}_s(0)=\pi$. In another case, the method of Dyson-Schwinger equation (DSE) defines a process-independent (PI) effective charge $\hat{\alpha}_{\rm PI}(Q)$~\cite{Binosi:2016nme, Cui:2019dwv}, which is derived from a gluon two-point function $D^{\rm PB}_{\mu\nu}(Q,\zeta)$ under the Pinch Technique and Background Field Method. After taking the solutions of DSE and lattice results, the PI effective charge $\alpha_{\rm PI}$ has a freezing value as $\hat{\alpha}_{\rm PI}(0)=(0.97\pm0.04)\pi$. Recent reviews offer a comparative analysis of the two distinct effective couplings, $\alpha^{g_1,{\rm LFHQCD}}_s(Q)$ and $\hat{\alpha}_{\rm PI}(Q)$, within the low-energy region~\cite{Deur:2023dzc,Brodsky:2024zev}. The finite predictions for effective coupling at a low $Q$-range under two different theories indicate that the interaction strengths may not significantly increase at certain energy scales under specific conditions, nor completely vanish but instead remain at a stable, non-zero level.

In high-energy region, based on the pQCD theory, the effective coupling $a^{g_1}_s(Q)$ can be written as an expansion of $a_s(\mu_r)=\alpha_s(\mu_r)/\pi$ as follows,
\begin{eqnarray}
a^{g_1}_s(Q)&=&\sum^{n}_{i=1}r^{\delta}_{i}(Q, \mu_r) a^{\delta, i}_s(\mu_r),
\label{ag1conv}
\end{eqnarray}
where the perturbative coefficients $r^{\delta}_i$ are calculated under a chosen renormalization scheme $\delta$ with the renormalization scale $\mu_r$. During the renormalization procedure, one chooses a typical renormalization scheme e.g. $\rm \overline{MS}$-scheme, and introduces the renormalization scale $\mu_r$ to cancel the divergence part in the loop momentum integrations. In principle, the renormalization scheme and scale dependence will vanish when the calculation order $n$ approaches infinity. Nevertheless, high-loop calculations are intricate, and perturbation computations performed to a finite order inevitably have the renormalization scheme and scale uncertainty. In this case, the renormalization scale $\mu_r$ is usually taken as an arbitrary value, e.g., typical momentum $Q$ and varying in $[Q/2,2Q]$ to estimate its uncertainty. This simple treatment results in an unphysical uncertainty i.e., the conventional renormalization scale uncertainty, which is however unnecessary and reduces the precision of the pQCD prediction, and violates the renormalization group invariance (RGI)~\cite{Petermann:1953wpa, GellMann:1954fq, Peterman:1978tb, Callan:1970yg, Symanzik:1970rt, Brodsky:1982gc}. In high-energy region, the relatively small $\alpha_s$ effectively mitigates the influence of this discrepancy, thereby preserving the predictive accuracy of the series. However, the substantial difference in the $\alpha_s$-value between the intermediate and low-energy domains inevitably reduces the accuracy of the smooth transition of $a^{g_1}_s(Q)$ from small to large scales, resulting in significant renormalization scale uncertainty~\cite{Yu:2021yvw}.

The occurrence of renormalization scale uncertainty in the fixed-order perturbative series arises from the mismatching between the perturbative coefficients and the values of $\alpha_s$. Based on the renormalization group invariance, the principle of maximum conformality approach (PMC)~\cite{Brodsky:2011ta, Mojaza:2012mf, Brodsky:2011ig, Brodsky:2012rj, Brodsky:2013vpa} is introduced to correctly determine the $\alpha_s$-value and eliminate this $\mu_r$-dependence with the help of RGE. That is, the PMC determines the magnitude of $\alpha_s$ by systematically using the non-conformal $\{\beta_i\}$-terms of the pQCD series, which then turns into a scheme-invariant conformal series and eliminates the $\mu_r$-dependence simultaneously. It has been proved that the PMC predictions are independent of the renormalization scheme and scale~\cite{Wu:2014iba, Wu:2018cmb, Wu:2019mky}.

The standard PMC procedures suggested in Refs.\cite{Brodsky:2011ta, Mojaza:2012mf, Brodsky:2011ig, Brodsky:2012rj, Brodsky:2013vpa} is a multi-scale setting approach, simply called as the PMCm approach, which uses the RGE and fixes the effective couplings for each order by absorbing different types of $\{\beta_i\}$-terms into the corresponding $\alpha_s$ via an order-by-order manner. Since the same type of $\{\beta_i\}$-term appearing in different orders, the corresponding PMCm scales at each order exhibit a perturbative nature. So even though the PMCm removes conventional scheme-and-scale ambiguities, apparently, the uncalculated/unknown higher-order terms will cause two kinds of residual scale dependence~\cite{Zheng:2013uja}, e.g. the last perturbative terms of all the PMCm scales are unknown, which is called as \textit{the first kind of residual scale dependence}; and since there is no $\{\beta_i\}$-terms to fix the PMCm scale for the highest-order terms of the pQCD series, its magnitude cannot be strictly fixed, which is called as \textit{the second kind of residual scale dependence}. In large $Q$-range, such residual scale dependence would be depressed by small magnitude of $\alpha_s$ and the perturbative nature of the PMC scales. But those two residual scale dependence could be sizable in low and intermediate $Q$-range.

To achieve a smooth transition for the effective coupling $a^{g_1}_s(Q)$ within whole energy region, one needs to fix a critical scale $Q_0$~\cite{Brodsky:2010ur, Deur:2014qfa, Deur:2016cxb}, indicating a transition between perturbative and non-perturbative domain, whose magnitude is fixed by requiring $\alpha_s(Q_0)$ from either side to be exactly the same and their derivatives from either side also to be exactly the same. Ref.\cite{Deur:2017cvd} has applied the PMCm approach to fix the effective coupling and $Q_0$, it has been found that the PMC series converges much faster than the conventional $\overline{\rm MS}$ pQCD series, which results in a significantly smaller uncertainty for the PMC prediction. Thus the PMC provides a determination of $\alpha^{g_{1}}_{s}(Q)$ compatible with the data and the conventional pQCD calculation, but without scheme-dependence and with significantly improved precision. However a ``self-consistency problem" has also been observed, e.g. the domain of applicability of the nonperturbative LFHQCD and the domain of applicability of the perturbative PMC predictions do not overlap if the $\overline{\rm MS}$-scheme is used as the auxiliary scheme, which is caused by the fact that the PMC scales at certain orders in the $\overline{\rm MS}$-scheme are in some cases smaller than the transition scale $Q_0$.

The PMC single-scale approach (PMCs)~\cite{Shen:2017pdu} and the PMC infinite-order scale-setting approach (${\rm PMC}_{\infty}$)~\cite{DiGiustino:2020fbk} have been suggested to suppress the residual scale dependence. The PMCs method uses a singlet PMC scale $Q_*$ to absorb all $\{\beta_i\}$-terms of the whole series. By using the PMCs, \textit{the second kind of residual scale dependence} is removed, the single effective PMC scale displays stability and convergence with increasing order in pQCD. Though the PMC prediction is scheme independent, a proper choice of scheme could have some subtle differences. Using the PMCs to deal with the BSR, Ref.\cite{Yu:2021yvw} shows that if further transforming the series from the $\rm \overline{MS}$-scheme into the physical $V$-scheme~\cite{Appelquist:1977tw, Fischler:1977yf, Peter:1996ig, Schroder:1998vy, Billoire:1979ih}, the above mentioned ``self-consistency problem" can be avoided. However due to perturbative nature of the PMCs scale and its perturbative series does not convergent enough, \textit{the first kind of residual scale dependence} arising from the unknown higher-order terms still exerts a significant influence on the theoretical forecasting of PMCs.

The ${\rm PMC}_{\infty}$ approach determines the coupling constant $\alpha_s(\mu_i)$ at each order by using the intrinsic conformality (iCF) of the series, which sets the scales by requiring that all the scale-dependent $\{\beta_i\}$-terms at each order vanish separately. This provides the required conformal coefficients via an order-by-order manner. The ${\rm PMC}_{\infty}$ is also a multi-scale-setting approach, but different from the PMCs, the $\alpha_s(\mu_i)$-values at low orders are determined after identifying all $\beta_0$-terms at each order, then the newly fixed PMC scales at each order are no-longer in perturbative form, thus it will not have the \textit{the first kind of residual scale dependence}. While, like the early PMCm, the last term of the resultant ${\rm PMC}_{\infty}$ series lacks enough information to determine its exact $\alpha_s$ value, e.g. it still has \textit{the second kind of residual scale dependence}. For a good convergent series of the pQCD approximant, \textit{the second kind of residual scale dependence} could be negligible when enough higher-order terms are known, thus improving the precision of the pQCD prediction.

A detailed comparison of various PMC scale-setting approaches can be found in Ref.\cite{Huang:2021hzr}, which shows that different scale-setting approaches may have its own advantages and one may choose a proper one to achieve a even more precise fixed-order prediction after removing conventional scheme-and-scale ambiguities. In this paper, we will adopt the PMC$_\infty$ method to eliminate such residual scale dependence with the attempt of further improving the precision of theoretical prediction.

The remaining parts of the paper are organized as follows. In Sec.\ref{sec2}, we present the main prescriptions for calculating $a_s^{g_1}(Q)$ via its perturbative relations to the $V$-scheme strong coupling by using both the PMC$_\infty$ and the PMCs procedures. In Sec.\ref{sec3}, we show the extracted values of $\alpha_s(M_z)$ through the effective coupling $a_s^{g_1}(Q)$, i.e., by meticulously comparing the pQCD predictions with either LFHQCD results or the experimental data. In Sec.\ref{sec4}, we summarize the results.

\section{calculation technology}
\label{sec2}

Taking the physical $V$-scheme~\cite{Appelquist:1977tw, Fischler:1977yf, Billoire:1979ih} strong coupling as the expansion basis, the effective coupling defined in Eq.(\ref{ag1conv}) can be written as
\begin{eqnarray}
a^{g_1}_s(Q)&=& \sum^{n}_{i=1}r_i^{\rm V}(Q,\mu_r)a^{{\rm V},i}_s(\mu_r),\nonumber\\
&=& r^{\rm V}_{1,0}a^{\rm V}_{s}(\mu_r) +\left(r^{\rm V}_{2,0}+\beta_0 r^{\rm V}_{2,1}\right)a^{\rm V,2}_{s}(\mu_r) \nonumber\\
& & +\left(r^{\rm V}_{3,0}+\beta_1 r^{\rm V}_{2,1}+2\beta_{0} r^{\rm V}_{3,1}+\beta^2_{0} r^{\rm V}_{3,2}\right) a^{\rm V,3}_{s}(\mu_r) \nonumber\\
& & +\Big(r^{\rm V}_{4,0}+\beta^{\rm V}_2 r^{\rm V}_{2,1} +2\beta_{1} r^{\rm V}_{3,1} + \frac{5}{2} \beta_0 \beta_1 r^{\rm V}_{3,2} +3\beta_0 r^{\rm V}_{4,1} \nonumber\\
& & +3 \beta^2_0 r^{\rm V}_{4,2} + \beta^3_0 r^{\rm V}_{4,3}\Big) a^{\rm V,4}_{s}(\mu_r) + {\cal O}(a^{\rm V,5}_{s}),
\label{ag1Vcon}
\end{eqnarray}
where $a^{V}_s(\mu_r) = {\alpha^{V}_s(\mu_r)} /{\pi}$ and the coefficients $r_{i,j}^{\rm V}$ up to $n=4$ can be derived by using the known coefficients calculated under the ${\overline{\rm MS}}$ scheme~\cite{Gorishnii:1985xm, Larin:1991tj, Baikov:2010je, Larin:2013yba, Baikov:2015tea, Blumlein:2016xcy}. This can be done with the help of the relation between the $V$-scheme and the ${\overline{\rm MS}}$-scheme RGEs~\cite{Kataev:2023sru, Kataev:2015yha}. For convenience, we put the transition procedures between those two schemes and the $V$-scheme coefficients $r^{\rm V}_{i,j}$ in the Appendix.

Based on the iCF assumption of standard PMC$_\infty$ method, the perturbative expansion of effective coupling $a^{g_1}_s(Q)$ exhibits a particular structure, which collects all the terms with the same conformal coefficients into the same conformal subset. For the $a^{g_1}_s(Q)$ expansion known up to $\mathcal{O}(a^{\rm V,4}_s)$, there are four different subsets,
\begin{eqnarray}
a^{g_1}_s(Q)&=&	\sum^{4}_{i=1}a^{g_1}_{s,i}(Q).
\label{ag1Vic}
\end{eqnarray}
Each subset $a^{g_1}_{s,i}(Q)$ with $i=(1,\cdots,4)$ collects together the same category of non-conformal terms and ensures the scheme independence of each subset via the commensurate scale relations among different orders~\cite{Brodsky:1994eh}. Each subset is scale invariant. To be specifically, the required four subsets have the following formalization
\begin{widetext}
\begin{eqnarray}
a^{g_1}_{1}(Q) &=& r^{\rm V}_{1,{\rm IC}}\bigg\{a^{\rm V,1}_s(\mu_r)+\beta_0 \ln{\mu^2_r\over\mu^{\rm V,2}_1}a^{\rm V,2}_s(\mu_r)+\left(\beta_1+\beta^2_{0}\ln{\mu^2_r\over\mu^{\rm V,2}_1}\right)\ln{\mu^{2}_r\over\mu^{\rm V,2}_1}a^{\rm V,3}_s(\mu_r)\nonumber\\&&
+\left[\beta^{\rm V}_2+\frac{5}{2} \beta_0 \beta_1\ln{\mu^2_r\over\mu^{\rm V,2}_1} +\beta^3_0 \bigg(\ln{\mu^2_r\over\mu^{\rm V,2}_1}\bigg)^2\right]\ln{\mu^2_r\over\mu^{\rm V,2}_1}a^{\rm V,4}_s(\mu_r)\bigg\}, \\
a^{g_1}_{2}(Q)&=& r^{\rm V}_{2,{\rm IC}}\left[a^{\rm V,2}_s(\mu_r)+2\beta_0\ln{\mu^2_r\over\mu^{\rm V,2}_2}a^{\rm V,3}_s(\mu_r)+\left(2\beta_{1}+3\beta^2_0\ln{\mu^2_r\over\mu^{\rm V,2}_2}\right)\ln{\mu^2_r\over\mu^{\rm V,2}_2}a^{\rm V,4}_s(\mu_r)\right],
\end{eqnarray}
\end{widetext}
\begin{eqnarray}
a^{g_1}_{3}(Q) &=& r^{\rm V}_{3,{\rm IC}} \left[a^{\rm V,3}_s(\mu_r)+3\beta_0\ln{\mu^2_r\over\mu^{\rm V,2}_3}a^{\rm V,4}_s(\mu_r) \right],\\
a^{g_1}_{4}(Q) &=& r^{\rm V}_{4,{\rm IC}} a^{\rm V,4}_s(\mu_r),
\end{eqnarray}
where $r^{\rm V}_{i,{\rm IC}}$ are the scale-independent iCF coefficients and $\mu^{\rm V}_i$ are the PMC scales. The coefficients $r^{\rm V}_{i,{\rm IC}}$ can be identified from the known $r^{\rm V}_i$ by using the standard PMC$_\infty$ procedure. For examples, at the leading order (LO) which is already conformal, $r^{\rm V}_{1,{\rm IC}}=r^{\rm V}_{1}$; and at the next-to leading order (NLO), $r^{\rm V}_{2,{\rm IC}}=r^{\rm V}_{2}(n_f=33/2)$, where fixing the quark flavor number $n_f=33/2$ makes the $\beta_0$-term vanish. For more high-order terms,
\begin{eqnarray}
r^{\rm V}_{3,{\rm IC}}&=&r^{\rm V}_{3}\bigg(n_f={33\over2}\bigg)-r^{\rm V}_{1,{\rm IC}}\overline{\beta}_1\ln{\mu^2_r\over\mu^{\rm V,2}_1},\\
r^{\rm V}_{4,{\rm IC}}&=&r^{\rm V}_{4}\bigg(n_f={33\over2}\bigg)
-2r^{\rm V}_{2,{\rm IC}}\overline{\beta}_1\ln{\mu^2_r\over\mu^{\rm V,2}_2}\nonumber\\&&
-r^{\rm V}_{1,{\rm IC}}\overline{\beta}^{\rm V}_2\ln{\mu^2_r\over\mu^{\rm V,2}_1},	
\end{eqnarray}
where $\overline{\beta}_1=\beta_1(n_f=33/2)$ and $\overline{\beta}^{\rm V}_2=\beta^{\rm V}_2(n_f=33/2)$. It has been proved that the conformal coefficients $r^{\rm V}_{i,{\rm IC}}\equiv r^{\rm V}_{i,0}$~\cite{Huang:2021hzr}. The PMC scales $\mu^{\rm V}_i$ are determined by requiring each subsets satisfying the condition of RGI, and its first three ones are
\begin{widetext}
\begin{eqnarray}
\ln\frac{\mu^2_r}{\mu^{\rm V,2}_1}&=& \frac{r^{\rm V}_{2}-r^{\rm V}_{2,{\rm IC}}}{r^{\rm V}_{1,{\rm IC}}\beta_0}, ~\label{q1}\\
\ln\frac{\mu_r^2}{\mu^{\rm V,2}_2}&=&\frac{r^{\rm V}_{3}-r^{\rm V}_{3,{\rm IC}}-r^{\rm V}_{1,\rm IC}\left[\beta_1+\beta^2_0\ln(\mu^2_r/\mu^{\rm V,2}_1)\right]\ln(\mu^2_r/\mu^{\rm V,2}_1)}{2r^{\rm V}_{2,{\rm IC}}\beta_0} ,~\label{q2} \\
\ln\frac{\mu_r^2}{\mu^{\rm V,2}_3}&=& \frac{r^{\rm V}_{4}-r^{\rm V}_{4,{\rm IC}}}{3r^{\rm V}_{3,{\rm IC}}\beta_0}
-\frac{r^{\rm V}_{2,{\rm IC}}\left[2\beta_1+3\beta^2_0\ln(\mu^2_r/\mu^{\rm V,2}_2)\right]\ln(\mu^2_r/\mu^{\rm V,2}_1)}{3r^{\rm V}_{3,{\rm IC}}\beta_0}
\nonumber\\&&
-\frac{r^{\rm V}_{1,{\rm IC}}\left[\beta^{\rm V}_2+\frac{5}{2}\beta_0\beta_1\ln(\mu^2_r/\mu^{\rm V,2}_1)+\beta^2_0\ln^2(\mu^2_r/\mu^{\rm V,2}_1)\right]\ln(\mu^2_r/\mu^{\rm V,2}_1)}{3r^{\rm V}_{3,{\rm IC}}\beta_0}.~\label{q3}
\end{eqnarray}
\end{widetext}

Then we can transform the initial series as the following series
\begin{eqnarray}
\centering
a^{g_1}_s(Q)|_{\rm PMC_\infty}&=&\sum^{4}_{i=1}r_{i,\rm IC}^{\rm V}(Q) \  a^{{\rm V},i}_s(\mu^{\rm V}_i).
\label{asg1pmc}
\end{eqnarray}
It is a conformal series, which incorporates accurate coupling values $a^{\rm V}_s(\mu_i)$ and exhibits scale invariance under the condition that all the RGE-involved $\{\beta_i\}$-terms vanish. This ensures that the series exhibits renormalization scale invariance and is free from the \textit{first kind of residual scale dependence}. Moreover, the $\rm PMC_\infty$ approach reveals the intrinsic conformality of the QCD perturbation series and achieves the ordered scale invariance, i.e., retaining scale invariance at any order. However, given the absence of higher-order information, the $\rm PMC_\infty$ last scale $\mu^{\rm V}_4$ remains uncertain, it still has the \textit{second kind of residual scale dependence}. Fortunately, this uncertainty associated with the $\rm N^4LO$-term is depressed by a high power of $\alpha^{\rm V}_s$ and is further mitigated by the excellent convergence properties of the series (\ref{asg1pmc}). If not specially stated, we will implicitly take $\mu^{\rm V}_4=\mu^{\rm V}_3$ for the subsequent discussions.

To compare the effects of the PMCs and the PMC$_\infty$ methods in eliminating the residual scale dependence due to unknown higher-order terms, we also apply the PMCs method to deal with the pQCD corrections to $a^{g_1}_s(Q)$. Using the standard PMCs procedures, the resultant PMC series is
\begin{eqnarray}
	a^{g_1}_s(Q)|_{\rm PMCs}=\sum^{4}_{i=1}r_{i,0}^{\rm V}(Q) \ a^{{\rm V},i}_s(Q^{\rm V}_*),
\label{pmcs0}
\end{eqnarray}
where by using the given N$^3$LO-level pQCD series, the single-scale $Q^{\rm V}_*$ can be fixed up to $\mathcal{O}(\alpha^3_s)$-level, e.g. the next-to-next-to-leading log (N$^2$LL) accuracy,
\begin{eqnarray}
\ln\frac{Q^{\rm V,2}_*}{Q^2}=S_{0}+S_{1}a^{\rm{V}}_s(Q^{\rm V}_*)+S_{2}a^{\rm{V},2}_s(Q^{\rm V}_*),
\label{singletscale}
\end{eqnarray}
whose expansion coefficients $S_i$ are
\begin{eqnarray}
S_0 &=& -{\hat{r}^{\rm V}_{2,1}\over \hat{r}^{\rm V}_{1,0}},  \\
S_1 &=& {2(\hat{r}^{\rm V}_{2,0}\hat{r}^{\rm V}_{2,1}-\hat{r}^{\rm V}_{1,0}\hat{r}^{\rm V}_{3,1})\over\hat{r}^{\rm V,2}_{1,0}} +{(\hat{r}^{\rm V,2}_{2,1}-\hat{r}^{\rm V}_{1,0}\hat{r}^{\rm V}_{3,2})\over \hat{r}^{\rm V,2}_{1,0}}\beta_0,
\end{eqnarray}
\begin{widetext}
\begin{eqnarray}
S_2&=&-\frac{\left(2 \hat{r}^{\rm V,3}_{2,1}-3
   \hat{r}^{\rm V}_{1,0} \hat{r}^{\rm V}_{3,2} \hat{r}^{\rm V}_{2,1}+\hat{r}^{\rm V,2}_{1,0}
   \hat{r}^{\rm V}_{4,3}\right)}{\hat{r}^{\rm V,3}_{1,0}}\beta _0^2 -\frac{\hat{r}^{\rm V}_{2,0} \left(5 \hat{r}^{\rm V,2}_{2,1}-2\hat{r}^{\rm V}_{1,0} \hat{r}^{\rm V}_{3,2}\right)
   +3 \hat{r}^{\rm V}_{1,0}\left(\hat{r}^{\rm V}_{1,0} \hat{r}^{\rm V}_{4,2}-2 \hat{r}^{\rm V}_{2,1}\hat{r}^{\rm V}_{3,1}\right)}{\hat{r}^{\rm V,3}_{1,0}}\beta_0 \nonumber\\
   &&    +\frac{3 \left(\hat{r}^{\rm V,2}_{2,1}-\hat{r}^{\rm V}_{1,0}
   \hat{r}^{\rm V}_{3,2}\right)}{2 \hat{r}^{\rm V,2}_{1,0}}\beta_1
   +\frac{4\hat{r}^{\rm V}_{2,0}\left(\hat{r}^{\rm V}_{1,0} \hat{r}^{\rm V}_{3,1}-\hat{r}^{\rm V}_{2,0} \hat{r}^{\rm V}_{2,1}\right)}{\hat{r}^{\rm V,3}_{1,0}} +\frac{3 \hat{r}^{\rm V}_{1,0} \left(\hat{r}^{\rm V}_{2,1}\hat{r}^{\rm V}_{3,0}-\hat{r}^{\rm V}_{1,0}\hat{r}^{\rm V}_{4,1}\right)}{\hat{r}^{\rm V,3}_{1,0}}.
\end{eqnarray}
\end{widetext}
The PMCs series (\ref{pmcs0}) do not have \textit{the second kind of residual scale dependence}, and because the single-scale $Q^{\rm V}_*$ is of perturbative nature, it still has \textit{the first kind of residual scale dependence}. We will use the the Pad$\acute{e}$ approximation approach (PAA)~\cite{Basdevant:1972fe} to estimate the magnitude of the N$^3$LL-term, i.e. the $\pm S_3 a^{\rm V,3}_s(Q^{\rm V}_*)$-term, which leads to $\Delta Q^{\rm V}_*$. Generally, such kind of residual scale dependence will be suffered from double suppression of the exponential-suppression and $\alpha_s$-suppression. However, if the series of the right-hand-side of Eq.(\ref{singletscale}) does not converge enough, because the error of $Q^{\rm V}_*$ works for the LO-terms of the series, \textit{the second kind of residual scale dependence} could be sizable.

\section{numerical results and discussions}
\label{sec3}

Using the RGE, the solution for the running behavior of $\alpha_s(\mu_r)$ can be expressed as an expansion over inverse powers of the logarithm $l_\Lambda=\ln{\mu_r^2/\Lambda^2}$. The universal $\alpha_s$ running behavior up to four-loop level is~\cite{Chetyrkin:1997sg, Brodsky:2011ta}
\begin{eqnarray}
a_s(\mu_r)&=&{1\over\beta_0l_\Lambda}\bigg\{1-{\beta_1\over\beta^2_0}{\ln{l_\Lambda}\over {l_\Lambda}}+{1\over\beta^2_0{l^2_\Lambda}}
\nonumber\\&&
\times\left[{\beta^2_1\over\beta^{2}_0}(\ln^2l_\Lambda
-\ln{l_\Lambda}-1)+{\beta_2\over\beta_0} \right]
\nonumber\\&&
+{1\over\beta^3_0l^3_\Lambda} \bigg[{\beta^{3}_1\over\beta^3_0}(-\ln^3{l_\Lambda}
+{5\over2}\ln^2{l_\Lambda}
+2\ln{l_\Lambda}
\nonumber\\&&
-{1\over2})
-3{\beta_1\beta_2\over\beta^2_0}\ln{l_\Lambda}+{\beta_3\over2\beta_0}\bigg] \bigg\},
\label{alphas}
\end{eqnarray}
where the asymptotic scale $\Lambda$ and $\{\beta_{i(\geq2)}\}$-functions are scheme-dependent. The $\Lambda$ plays a pivotal role in defining the boundary for the truncation of infinite integrations encountered in pQCD calculations, it can be associated with the typical hadron size and its value is not predicted by the QCD theory but must be extracted from a measurement of $\alpha_s$ at a given reference scale. Typically, the value of $\Lambda$ is determined by fixing a particular $\alpha_s$ at a standard scale, such as the mass of the $Z$ boson ($M_Z$), within a specified renormalization scheme.

\begin{figure}[htbp]
\centering
\includegraphics[width=0.45\textwidth]{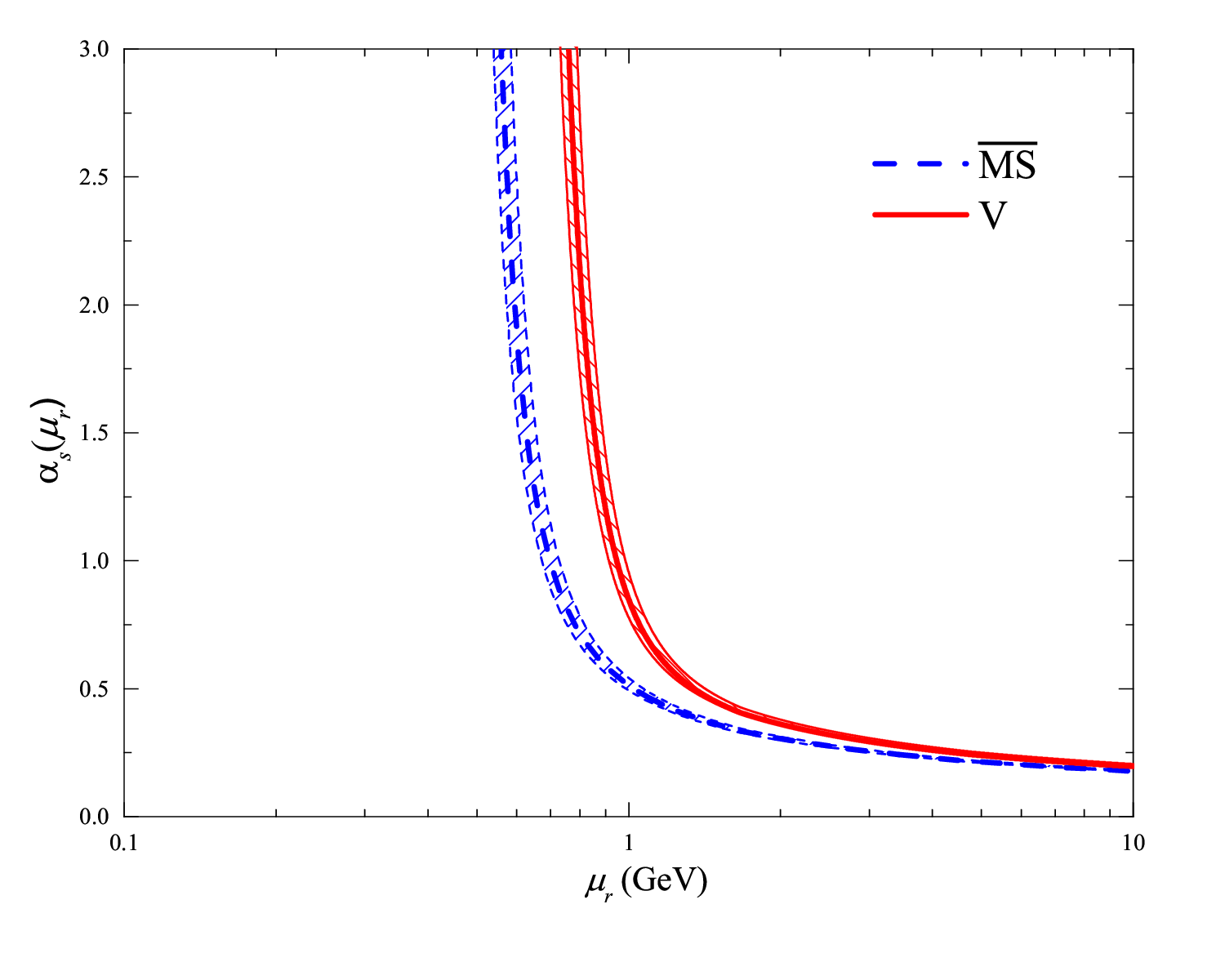}
\caption{The four-loop $\alpha_s(\mu_r)$ running behavior under ${\rm \overline{\rm MS}}$-scheme and $V$-scheme, respectively. The blue band shows the running behavior with $\Lambda^{n_f=5}_{\rm \overline{MS}}=208\pm10$ MeV under the $\rm \overline{\rm MS}$-scheme and the red band shows the running behavior of $\alpha_s$ with $\Lambda^{n_f=5}_{\rm V}=285\pm14$ MeV under the $V$-scheme. }
\label{pic1}
\end{figure}

To do numerical calculation, we adopt the PDG averaged value $\alpha_s(M_Z) =0.1180\pm0.0009$~\cite{ParticleDataGroup:2024cfk} to fix $\alpha_s$ running behavior, which are shown in Fig.\ref{pic1}. At the mass scale $M_Z=91.1876$ GeV and the quark flavor $n_f=5$, we obtain $\Lambda^{n_f=5}_{\rm \overline{MS}}=208\pm10$ MeV under the $\overline{\rm MS}$-scheme, and $\Lambda^{n_f=5}_{\rm V}=285\pm14$ MeV under the $V$-scheme.

\subsection{PQCD corrections of $\alpha^{g_1}_{s}$ up to $\rm N^3LO$-level}
\label{secA}

\begin{table*}[htb]
\centering
\begin{tabular}{  c c  c  c c c c}
\hline
 & ~~~   &~~LO~~~   &~~~NLO~~~  &~~~$\rm N^2LO$~~~  &~~~$\rm N^3LO$~~~ &~~~$\rm total$~~~\\
\hline
&Conv.    &$0.129^{+0.141}_{-0.046}$  &$0.024^{-0.088}_{+0.013}$ &$-0.003^{-0.134}_{+0.021}$ &$0.011^{+0.198}_{+0.003}$ &$0.161^{+0.117}_{-0.009}$ \\
&PMC$_{\rm \infty}$   &0.108     &0.020   &0.029   &0.005  &0.162\\
&PMCs                &0.114    &0.035   &0.027   &0.005  &0.181\\
\hline
\end{tabular}
\caption{Results for effective coupling $a^{g_1}_s(m_c)$ at each order (LO, NLO, $\rm N^2LO$ and $\rm N^3LO$) under the conventional (Conv.), the PMC$_\infty$ and PMCs methods, respectively. The conventional predictions are highly scale dependent, whose central values are for $a^{g_1}_{s}(m_c)|_{\rm Conv.}$ with $\mu_r=m_c$ and the uncertainties are for $\mu_r\in[1, 4]$ GeV.}
\label{table1}
\end{table*}

Under $\overline{\rm MS}$-scheme, Refs.~\cite{Gorishnii:1985xm,Larin:1991tj,Baikov:2010je,Larin:2013yba,Baikov:2015tea} have offered the analytic perturbation corrections of the effective coupling $a^{g_1}_{s}(Q)$ up to four-loop. In our present calculations, we have completed the transformation of Eq.(\ref{ag1Vcon}) into the $V$-scheme and the heavy quark mass effects are also considered in the perturbation coefficient $r_2$~\cite{Blumlein:2016xcy}, when $m_c=1.67$ GeV and $m_b=4.78$ GeV are taken. At the typical momentum $Q=m_c$, the effective coupling $a^{g_1}_s(m_c)$ takes the form
\begin{eqnarray}
a^{g_1}_{s}(m_c)|_{\rm Conv.} &=& a^{\rm V}_s (m_c)+1.429 a^{\rm V,2}_s (m_c)-1.586 \nonumber\\&&\times a^{\rm V,3}_s(m_c)
+40.442 a^{\rm V,4}_s (m_c),
\label{ag1c}
\end{eqnarray}
in which $\mu_r$ is chosen to be $m_c$. The results for different choices of $\mu_r$ can be derived with the help of RGE. Considering the error by taking $\mu_r\in[1, 4]$ GeV, contributions of the LO-terms, the NLO-terms, the $\rm N^2LO$-terms and the $\rm N^3LO$-terms for the effective coupling $a^{g_1}_s(m_c)$ are shown in Table.\ref{table1}. Table.\ref{table1} gives the results for conventional (Conv.), the PMCs and PMC$_\infty$ methods, respectively. The net value for $a^{g_1}_s(m_c)$ under the conventional scale setting method is $a^{g_1}_{s}(m_c)|_{\rm Conv.}=0.161^{+0.117}_{-0.009}$, which shows the scale error $\sim(^{+73\%}_{-6\%})$.

After applying the PMC$_\infty$ method, the corresponding IC perturbative series is
\begin{eqnarray}
a^{g_1}_{s}(m_c)|_{\rm PMC_{\infty}} &=& a^{\rm V}_s (\mu^{\rm V}_1)+2.741 a^{\rm V,2}_s (\mu^{\rm V}_2)+ 18.838
\nonumber\\&&\times a^{\rm V,3}_s(\mu^{\rm V}_3)
+27.024 a^{\rm V,4}_s (\mu^{\rm V}_4),
\label{ag1inf}
\end{eqnarray}
where the PMC scales $\mu^{\rm V}_i$ are functions of the chosen momentum $m_c$ and independent of the renormalization scale $\mu_r$. From the Eqs.(\ref{q1},\ref{q2},\ref{q3}), the three $\rm PMC_{\infty}$ scales of Eq.(\ref{ag1inf}) are $\mu^{\rm V}_1(m_c)=2.236$ GeV, $\mu^{\rm V}_2(m_c)=3.728$ GeV, and $\mu^{\rm V}_3(m_c)=1.988$ GeV, respectively. Additionally, we also give the PMCs prediction with the single-scale $Q^{\rm V}_*$ for comparison.
\begin{eqnarray}
a^{g_1}_{s}(m_c)|_{\rm PMCs} &=& a^{\rm V}_s (Q^{\rm V}_*)+2.741 a^{\rm V,2}_s (Q^{\rm V}_*)+ 18.838
\nonumber\\&&\times a^{\rm V,3}_s(Q^{\rm V}_*)
+27.024 a^{\rm V,4}_s (Q^{\rm V}_*),
\label{pmcs}
\end{eqnarray}
where $Q^{\rm V}_*(m_c)=2.042$ GeV.

Table.\ref{table1} shows the relative magnitudes between the LO-term, the NLO-term, the $\rm N^2LO$-term and the $\rm N^3LO$-term after applying the PMCs and PMC$_\infty$ approaches. It indicates that the PMC predictions are independent of the choice of $\mu_r$. The ratios of the magnitudes of N$^{i}$LO-terms over that of the LO-terms are introduced to explain the convergence of pQCD series under various scale-setting approaches, e.g., for conventional approach, we have
\begin{align*}
&1^{+1.093}_{-0.357}:0.186^{-0.683}_{+0.101}:-0.023^{-1.039}_{+0.163}:0.085^{+1.535}_{+0.023}
\end{align*}
and for PMC$_\infty$ and PMCs approaches, we have
\begin{align*}
&1:0.185:0.269:0.046,&\\
&1:0.307:0.237:0.044.&
\end{align*}
It is noted that since the magnitude of $\alpha_s$-value $\sim0.4$, the convergence of pQCD series for $a^{g_1}_s(m_c)$ is weak in the charm energy region. For conventional series with $\mu_r=m_c$, the magnitude of N$^3$LO-terms is even larger than that of the N$^2$LO-terms. After comparing the perturbative coefficients of $a^{g_1}_s(m_c)$ before and after applying the PMC, it is found that the PMC conformal series can improve the convergence of the perturbative series, even in charm energy region. It is because the PMC effective couplings absorb all the divergent non-conformal $\{\beta_i\}$-terms and demonstrates an intrinsic perturbative convergence of the series.

\begin{figure}[htbp]
\centering
\includegraphics[width=0.45\textwidth]{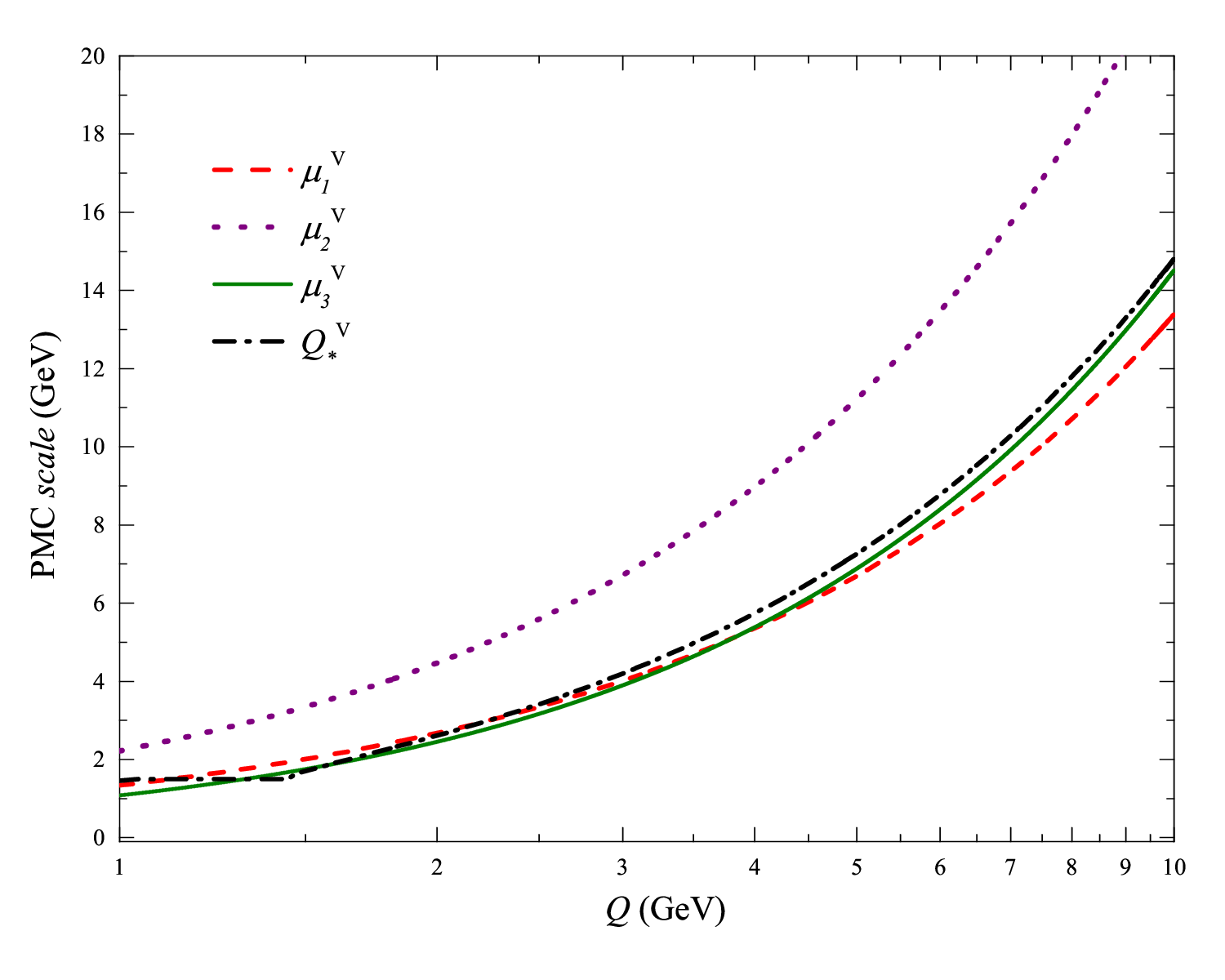}
\caption{The PMC scales for the $\alpha^{g_1}_{s}$ expansion over $V$-scheme strong coupling versus the momentum transfer $Q$, respectively. The red dashed line, the purple dotted line and green solid line present the first three $\rm PMC_\infty$ scales $\mu^{\rm V}_i$, respectively. The PMCs scale $Q^{\rm V}_*$ is presented a black dot-dashed line.}
\label{pic2}
\end{figure}

Fig.\ref{pic2} illustrates the variation of PMC scales for the $\alpha^{g_1}_{s}$ expansion over $V$-scheme strong coupling versus the momentum transfer $Q$. Fig.\ref{pic2} shows that all the PMC scales will increase with the increment of $Q$. And such rapid escalation of $\mu^{\rm V}_i$ also explains why our presently derived PMC scales under the $V$-scheme can solve the previously encountered ``self-consistency problem"~\cite{Deur:2017cvd} within the region $Q\gtrsim 1$ GeV.

For the PMCs procedure, the single-scale $Q^{\rm V}_*$ is determined by absorbing all types of $\{\beta_i\}$-terms into an overall effective $\alpha_s$, and the unknown higher-order $\{\beta_i\}$-terms directly leads to \textit{the first kind of residual scale dependence}. Using Eq.(\ref{singletscale}), we obtain $Q^{\rm V}_*(Q=m_c)=2.042$ GeV with the coefficients $S_0=0.583$, $S_1=5.607$ and $S_2=-63.362$. We then adopt the PAA approach to estimate the N$^3$LL-level perturbative coefficient $S_3$ for $\ln({Q^{\rm V,2}_*}/{Q^2})$ at $Q=m_c$, and by using the [0/1]-type generating function~\cite{Du:2018dma} we obtain
\begin{eqnarray}
S^{\rm PAA}_3 &=& {S^2_2\over S_1}. \label{PAA01}
\end{eqnarray}
By taking $\pm S^{\rm PAA}_3 a^{\rm V,3}_s(Q^{\rm V}_*)$ as an estimation of \textit{the first kind of residual scale dependence}, we obtain $Q^{\rm V}_*(Q=m_c)=2.042^{+0.917}_{+0.169}$ GeV, which leads to $a^{g_1}_s(m_c)|_{\rm PMCs}=0.181^{-0.045}_{-0.012}\ (\Delta Q^{\rm V}_*)$, indicating that the residual scale dependence for PMCs is about $\left(^{-24.9\%}_{-6.6\%}\right)$.

Different from the PMCs scale $Q^{\rm V}_*$, the $\rm PMC_\infty$ effective scales are fixed order-by-order and free of perturbative nature. Its uncertainty comes from the unknown last PMC scale $\mu^{\rm V}_4$ in Eq.(\ref{ag1inf}), which needs a higher-order $r^{\rm V}_5$-term to determine. Thus the $\rm PMC_\infty$ does not have \textit{the first kind of residual scale dependence}, but has \textit{the second kind of residual scale dependence}. As an estimation of \textit{the second kind of residual scale dependence} for $Q=m_c$, we take $\mu^{\rm V}_4(Q)\equiv\mu^{\rm V}_3(Q)$ and vary it from $Q$ to $2Q$ to estimate its uncertainty, which gives $\Delta a^{g_1}_s(Q=m_c)|_{\rm PMC_\infty}=\pm0.003(\Delta\mu^{\rm V}_4)$.
As the decoupling mass corrections have calculated up to $\mathcal{O}(\alpha^2_s)$-level~\cite{Blumlein:2016xcy}, we make a simple discussion on the charm and bottom quark mass effects on the effective coupling $a^{g_1}_{s}(Q)$. For the purpose, we set the momentum transfer $Q=m_{c,0}$ with $m_{c,0}=1.67$ GeV to do the discussion. The results for other momentum transfers can be done via a similar way. And the uncertainties for $a^{g_1}_{s}(Q=m_{c,0})$ are estimated by taking $\Delta m_c=\pm 0.5$ GeV and $\Delta m_b=\pm 1.0$ GeV. We then obtain $a^{g_1}_{s}(m_{c,0})|_{\rm Conv.}=0.1607^{+0.0268}_{-0.0025}(\Delta m_c)^{+0.0003}_{-0.0002}(\Delta m_b)$, which shows that $\Delta m_{c,b}$-errors $\sim(^{+11.8\%}_{-1.1\%})$. Here the errors for $a^{g_1}_{s}(m_{c,0})|_{\rm Conv.}$ is a combination of the mass corrections and renormalization scale dependence, since $\mu_r$ is fixed to $m_c$ for the scale-dependent conventional series. On the other hand, the PMC$_\infty$ prediction is, $a^{g_1}_{s}(m_{c,0})|_{\rm PMC_\infty}=0.1616^{+0.0002}_{-0.0001}(\Delta m_c)\pm{0.0001}(\Delta m_b)$ with the $\Delta m_{c,b}$-errors $\sim\pm0.1\%$; And the PMCs result is $a^{g_1}_{s}(m_{c,0})|_{\rm PMCs}=0.1813^{-0.0001}_{+0.0002}(\Delta m_c)^{-0.0001}_{+0.0002}(\Delta m_b)$ with $\Delta m_{c,b}$-errors $\sim\mp{0.1\%}$. Both indicate that the heavy quark mass effects are negligible for the PMC series.

Above numerical results show that the PMCs and PMC$_\infty$ series are consistent with each other with reasonable errors, which also show that due to the convergent behavior occurs at the N$^3$LO-level, the residual scale errors of the PMCs and PMC$_\infty$ series are smaller than the conventional scale error of the initial pQCD series. This shows the importance of a proper scale-setting approach. Moreover, the convergent behavior of the PMC$_\infty$ series for $a^{g_1}_s(Q)$ leads to a much smaller residual scale dependence, e.g. \textit{the second kind of residual scale dependence} is much smaller than \textit{the first kind of residual scale dependence} of PMCs series. In fact, even by varying $\mu^{\rm V}_4(Q=m_c)$ up to $8Q$, the magnitude of $\Delta a^{g_1}_s(Q=m_c)$ will only be changed to $0.004$. Thus, a precise prediction for the properties of $a^{g_1}_s(Q)$ may be achieved by using PMC$_\infty$ series. So in the following, we will utilize the PMC$_\infty$ approach to do the discussions.

\subsection{The running behavior of $\alpha^{g_1}_{s}(Q)$ at all scales}

There is divergent behavior of $\alpha_s$ when energy scale near the asymptotic scale $\Lambda$. Many low-energy models have been suggested to explain the infrared behavior of $\alpha_s$, cf. the reviews~\cite{Prosperi:2006hx, Deur:2016tte}. In this subsection, we utilize the LFHQCD method for $a^{g_1}_s$ low-energy behavior and determine its running behavior at all scales in combination with the above perturbative behavior. The LFHQCD transforms the modified ${\rm AdS}_5$ coupling into momentum space and gives
\begin{eqnarray}
a^{ g_{1}, {\rm LFHQCD}}_{s}(Q) = e^{-Q^2/4\kappa^2},
\label{ag1LFH}
\end{eqnarray}
where the free parameter $\kappa$ can be derived from the light meson and baryon spectroscopy, which gives $\kappa=0.523\pm0.024$ GeV~\cite{Brodsky:2016yod}. In infrared region, the LFHQCD model gives $a^{ g_{1}, {\rm LFHQCD}}_{s}(0)=1$. To achieve a smooth scale-running behavior of $\alpha^{g_1}_s(Q)$ across the whole $Q$-range, one can match the low-energy LFHQCD model with its behavior in perturbative behavior by fixing proper values for the critical scale $Q_0$ and $\kappa$. Their values can be fixed by two criterions, e.g. 1) the value of $\alpha^{g_1}_s(Q_0)$ calculated from both sides is exactly the same, and 2) the derivatives of $a^{g_1}_s(Q)$ at the $Q=Q_0$ point calculated from both sides is exactly the same. Here $Q_0$ serves as a threshold of perturbative and non-perturbative separation, for the case of $Q>Q_0$, $a^{g_1}_s(Q)$ is calculated by using Eqs.(\ref{ag1c},\ref{ag1inf},\ref{pmcs}), respectively; and for the case of $Q<Q_0$, $a^{g_1}_s(Q)$ is calculated by using Eq.(\ref{ag1LFH}).

\begin{figure}[htb]
\centering
\includegraphics[width=0.45\textwidth]{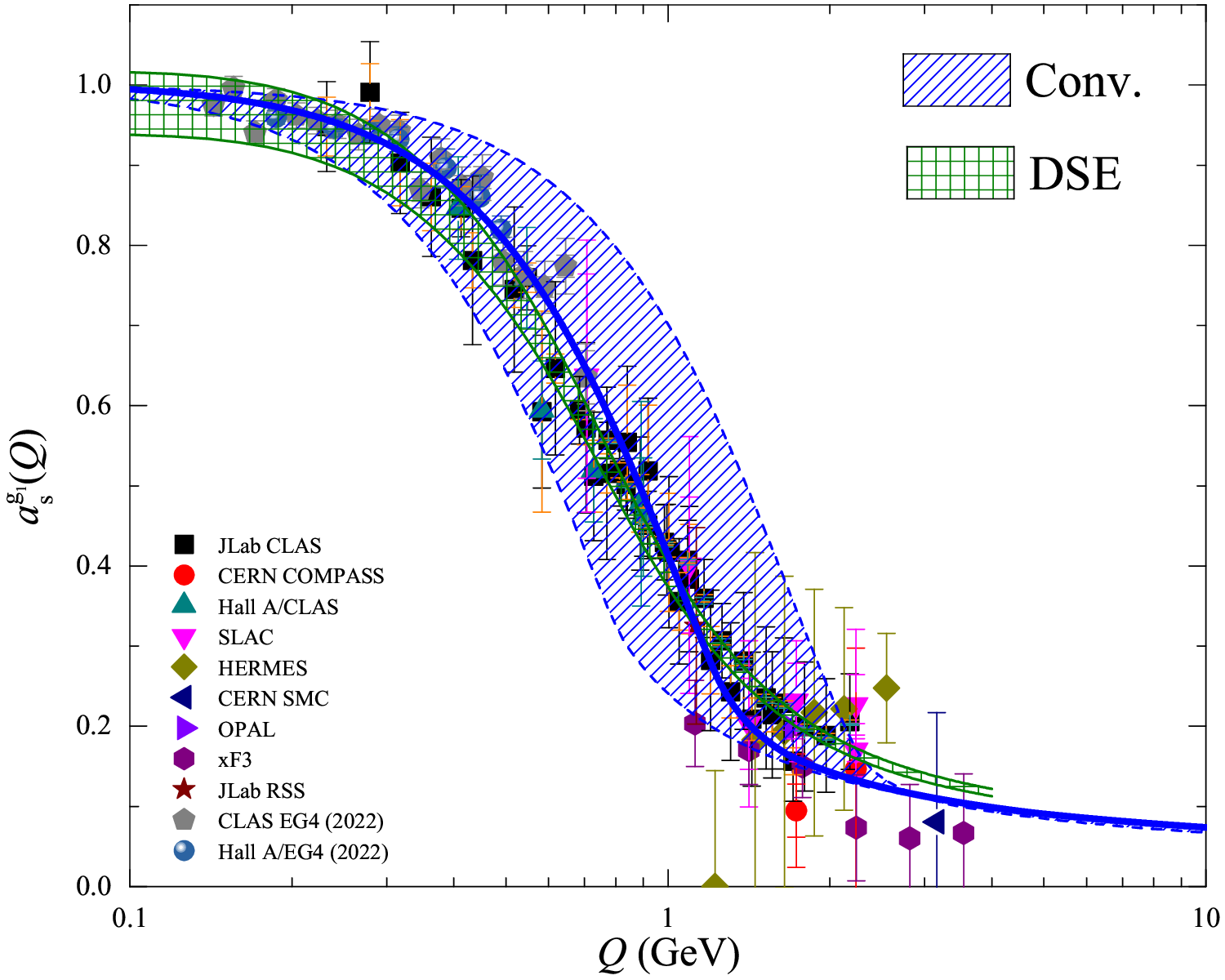}
\caption{The matching of the LFHQCD model of $a^{g_1}_s(Q)$ with its pQCD series up to ${\rm N^3LO}$-level QCD corrections, under conventional scale-setting method. The central blue line represents the conventional pQCD series when $\mu_r=Q$, and the blue band is obtained after considering the renormalization scale uncertainty with $\mu_r\in[Q,2Q]$. As a comparison, the DSE prediction~\cite{Binosi:2016nme, Cui:2019dwv} is also given in the green band.}
\label{pic3c}
\end{figure}

\begin{figure}[htb]
\centering
\includegraphics[width=0.45\textwidth]{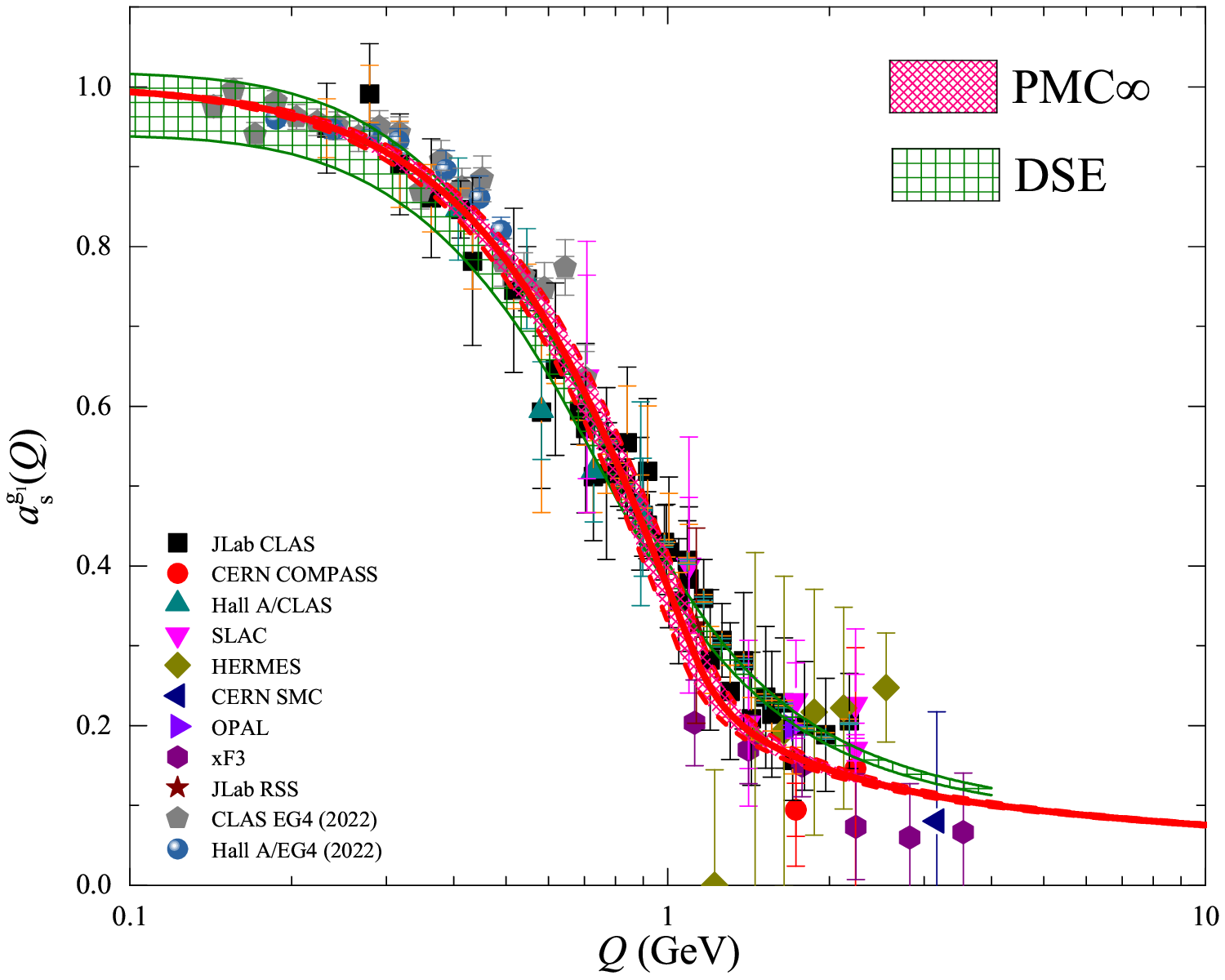}
\caption{The matching of the LFHQCD model of $a^{g_1}_s(Q)$ with its pQCD series up to ${\rm N^3LO}$-level QCD corrections, under PMC$_\infty$ approach. The red band represents a combination of the residual scale dependence estimated by using $\Delta\mu_4\in[Q, 2Q]$ and the error from $\Delta\Lambda^{n_f=5}_{\rm V}=14$ MeV. As a comparison, the DSE prediction~\cite{Binosi:2016nme, Cui:2019dwv} is also given in the green band.}
\label{pic3s}
\end{figure}

Fig.\ref{pic3c} and Fig.\ref{pic3s} show the matching of the low-energy LFHQCD model of $a^{g_1}_s(Q)$ with its pQCD series up to ${\rm N^3LO}$-level QCD corrections under conventional and PMC$_\infty$ approaches, respectively, which can be adopted to fix the parameters $Q_0$ and $\kappa$. The available experiment data points are also presented~\cite{Ackerstaff:1997ws, Alexakhin:2006oza, Anthony:1993uf, Abe:1997cx, Anthony:1999py, Adeva:1998vv, Ackerstaff:1998yj, Brodsky:2002nb, Gross:1969jf, Kim:1998kia, Adeva:1998vv,Deur:2004ti, Deur:2005cf, Deur:2008ej, Deur:2014vea, Deur:2022msf} in Fig.\ref{pic3c} and Fig.\ref{pic3s}. In small $Q$-region, the pQCD series of conventional scale-setting approach is highly scale dependent. In Fig.\ref{pic3c}, the blue solid line is for the case of $\mu_r=Q$, which leads to $\kappa=0.529$ GeV and $Q_0=1.205$ GeV. And for each $\mu_r$ within the range of $[Q, 2Q]$, one can achieve a smooth transition from the low to large energy region, with $\kappa=0.529^{+0.311}_{-0.153}$ GeV and $Q_0=1.205^{+0.976}_{-0.444}$ GeV, which show the $\mu_r$-uncertainty $\sim(^{+58.8\%}_{-28.9\%})$ in $\kappa$ and $\sim(^{+81.0\%}_{-36.8\%})$ in $Q_0$. After applying the PMC$_\infty$ approach to eliminate such scale dependence, Fig.\ref{pic3s} shows a more precise matching of the LFHQCD model with the PMC pQCD series can be achieved. In Fig.\ref{pic3s}, the central red solid line represents the matching with the PMC$_\infty$ prediction $a^{g_1}_s|_{\rm PMC_\infty}(Q)$, and the small red band shows the combined errors caused by the estimated residual scale dependence for $\Delta\mu^{\rm V}_4\in[Q,2Q]$ and the error from $\Delta\Lambda^{n_f=5}_{\rm V}=14$ MeV. As for the two input parameters, we have $\kappa=0.501^{+0.030}_{-0.028}$ GeV whose errors $\sim(^{+6.0\%}_{-5.6\%})$, and $Q_0=1.130^{+0.066}_{-0.059}$ GeV whose errors $\sim(^{+5.8\%}_{-5.2\%})$. Because of a faster increase of the PMC scales versus $Q$ for the $V$-scheme coupling, all the PMC scales are larger than $Q_0$, i.e. $\mu^{\rm V}_1(Q_0)$=$1.513^{+0.088}_{-0.078}$ GeV, $\mu^{\rm V}_2(Q_0)$=$2.516^{+0.147}_{-0.132}$ GeV, $\mu^{\rm V}_3(Q_0)$=$1.253^{+0.087}_{-0.078}$ GeV at $Q_0=1.130^{+0.066}_{-0.059}~{\rm GeV}$. This confirms our previous observation that by applying the PMC, the ``self-consistency problem" can be avoided. A comparison of Fig.\ref{pic3c} and Fig.\ref{pic3s} indicates that the PMC approach does improve the precision of theoretical predictions for $a^{g_1}_s(Q)$. It is found that the conventional line of $a^{g_1}_s(Q)$ under the choice of $\mu_r=Q$ is within the error band of PMC prediction.

The quality parameter $\chi^2/d.o.f$ (where the symbol ``d.o.f" is the short notation of the degrees of freedom) has been introduced in the literature to show the fitness of the theoretical predictions with the data. The $\chi^2/d.o.f$ indicates the quality of fit $\chi^2$ over the sum of the number of experiment data point $n$ minus the number of input parameters (for the present case, $\kappa$ and $Q_0$), which gives
\begin{eqnarray}
\chi^2/d.o.f=\frac{1}{n-2}\sum^{n}_{m=1} \frac{\left[a^{g_1, {\rm exp.}}_{s}(Q_m)-a^{g_1, {\rm the.}}_{g1}(Q_m)\right]^2} {\sigma^2_{m, {\rm exp.}}+\sigma^2_{m, {\rm the.}}},
\end{eqnarray}
where ``exp." stands for the experimental value, and each data point $Q_m$ with the experimental error $\sigma^2_{m, {\rm exp}}$, which includes both statistical error and systematic error. The symbol ``the." stands for the theoretical prediction, and $\sigma^2_{m, {\rm the.}}$ is its uncertainty. When doing the calculations of $\chi^2/d.o.f$, we take $n=79$ which includes the latest JLab experiment data~\cite{Deur:2004ti, Deur:2005cf, Deur:2008ej, Deur:2014vea, Deur:2022msf}. The conventional prediction is highly scale dependent, and for definiteness, we take the results for $\mu_r=Q$ to calculate its $\chi^2/d.o.f$, which is $\simeq 0.1$~\footnote{Its $p$-value is close to $100\%$, which is however due to a much larger theoretical error as shown by Fig.\ref{pic3c}.}. As for the PMC$_\infty$ prediction, we obtain $\chi^2/d.o.f \simeq0.6$. The smaller $\chi^2/d.o.f$-value obtained by conventional pQCD series is mainly due to large $\mu_r$-dependence in low and medium $Q$-region. Different form the conventional results, the PMC predictions are $\mu_r$-independent, and the good fit of the PMC to the experimental data further indicates the two parameters ($\kappa$ and $Q_0$) in PMC are reasonable and reliable, which corresponds to a $p$-value~\cite{ParticleDataGroup:2024cfk} $\sim 99\%$.

As a final remark, we also add the DSE model for a process-independent effective charge $\hat{a}_{\rm PI}(Q)$ in Fig.\ref{pic3c} and Fig.\ref{pic3s}. Inspired by the results of the strong coupling from the ghost-gluon vertex by lattice QCD~\cite{Zafeiropoulos:2019flq}, the DSE model connects the pinch technique gluon self-energy and the ghost-gluon vertex, and produces a modified strong coupling through a unique gluon dressing function:
\begin{eqnarray}
\hat{a}_{\rm PI}(Q)={\alpha_0\over\pi}{D(Q)\over\mathcal{D}(Q)} \left[\frac{F(Q;\mu_r)/F(0;\mu_r)}{1-L(Q;\mu_r)F(Q;\mu_r)}\right],
\end{eqnarray}
where the ratio of ${D(Q)/\mathcal{D}(Q)}$ is a RGI function, which defined in Refs.\cite{Binosi:2016nme, Cui:2019dwv}. The function $F(Q;\mu_r)$ is a redefined dressing function for the ghost propagator and $L(Q;\mu_r)$ is a longitudinal part of the gluon-ghost vacuum polarization, which are relay on an input $\mu_r=3.6$ GeV. The parameter $\alpha_0=0.97(4)\pi$, indicating the freezing value of the effective charge, is determined by the lattice calculation. Fig.\ref{pic3s} shows the $\hat{a}_{\rm PI}(Q)$ and PMC prediction $a^{g_1}_s(Q)$ are compatible in the small $Q$-region below the critical scale $Q_0$, whereas above the threshold, the PMC prediction show better agreement with the data.

\subsection{Determination of $\alpha_s(M_Z)$-value through the precise coupling $a^{g_1}_s(Q)$}

\begin{table*}[htb]
\centering
\begin{tabular}{  c c  c  c c c c c c}
\hline
 & ~~~   &~$Q_0$ (GeV)~~   &~~~$\Lambda^{n_f=3}_{\rm QCD}$(MeV)~~~   &~~~$\alpha_s(M_Z)$~~~  \\
\hline
&Conv.($\mu_r=Q$)                     &$1.1920^{+0.0541}_{-0.0542}$$^{+0.1684}_{-0.1859}$     &$424\pm20^{-154}_{+146}$     &$0.1177\pm{0.0011}^{-0.0095}_{+0.0077}$\\
&PMC$_\infty$  &$1.1782^{+0.0511}_{-0.0512}$$^{+0.0013}_{+0.0002}$     &$449\pm22\mp11$     &$0.1191\pm{0.0012}\mp0.0006$\\
\hline
\end{tabular}
\caption{The fitting parameters $Q_0$ and $\Lambda_{\rm QCD}$ and $\alpha_s(M_Z)$-value under the choise of $\kappa=0.523\pm0.024$ GeV. The first and second errors for $Q_0$ and $\alpha_s(M_Z)$ under conventional method are from the uncertainties of $\Delta\kappa=\pm0.024$ GeV and $\mu_r\in[Q/2,2Q]$, respectively. The first and second errors of the $Q_0$ and $\alpha_s(M_Z)$ under PMC$_\infty$ method are from the uncertainties of $\Delta\kappa=\pm0.024$ GeV and $\mu_4\in[Q,2Q]$, respectively.}
\label{table3}
\end{table*}

The strong coupling $\alpha_s$ is not an observable. So, the determination of the $\alpha_s$-value is usually through the comparison of some observables that can be precisely calculated and be precisely measured, e.g., the $e^+e^-$ hadronic cross section and the $\tau$, $W$ and $Z$ branching fractions to hadrons, and the hadronic decay width of heavy quarkonium~\cite{Proceedings:2019pra, dEnterria:2022hzv, Shen:2023qgz}.

In this subsection, we endeavor to harness the precision of PMC predictions for $a^{g_1}_s$ to meticulously determine the $\alpha_s(M_Z)$-value. And in the preceding alignment of LFHQCD and PMC predictions for a $a^{g_1}_s(Q)$, it is noteworthy that $\Lambda_{\rm QCD}$, being another parameter in the realm of nonperturbative physics, can likewise be determined through this matching procedure. For the purpose, by inversely taking the mass parameter $\kappa=0.523\pm0.024$ GeV derived from the calculation of hadron spectrum~\cite{Brodsky:2016yod} as an input, the other two matching parameters ($Q_0$ and $\Lambda_{\rm QCD}$) are shown in Table.\ref{table3}. For the case of $\mu_r=Q$, the conventional scale-setting method yields $Q_0=1.1920^{+0.0541}_{-0.0542}(\Delta\kappa)^{+0.1684}_{-0.1859}(\rm the.)$ GeV and $\Lambda^{n_f=3}_{\rm V}=424\pm20(\Delta\kappa)^{-154}_{+146}(\rm the.)$ MeV, where the first error ``$\Delta\kappa$" stems from $\Delta\kappa=\pm0.024$ GeV and the second error ``$\rm the.$" is from $\mu_r\in[Q/2,2Q]$. The PMC$_\infty$ method obtains $Q_0=1.1782^{+0.0511}_{-0.0512}(\Delta\kappa)~^{+0.0013}_{+0.0002}(\rm the.)$ GeV and $\Lambda^{n_f=3}_{\rm V}=449\pm22(\Delta\kappa)\mp{+11}(\rm the.)$ MeV, where the second error ``$\rm the.$" is from $\Delta\mu^{\rm V}_{4}\in[Q,2Q]$.

We then calculate the $\alpha_s$-value by considering the 4-loop running equation for the strong coupling $a_s(Q)$ given in Eq.(\ref{alphas}). And the input QCD scale $\Lambda^{n_f=5}_{\rm \overline{MS}}$ can be obtained through the asymptotic scale transformation relation between $V$ and $\rm \overline{MS}$ schemes in Appendix. The conventional prediction is
\begin{equation}
\alpha_s(M_Z)=0.1177\pm{0.0011}(\Delta\kappa)^{-0.0095}_{+0.0077}({\rm the.}),
\end{equation}
which shows the $\mu_r$-uncertainty $\sim(_{+6.5\%}^{-8.1\%})$. The PMC$_\infty$ prediction is
\begin{equation}
\alpha_s(M_Z)=0.1191\pm{0.0012}(\Delta\kappa)\mp0.0006({\rm the.}).
\end{equation}
whose net error is ignorable $\sim\mp0.5\%$. The PMC$_\infty$ prediction is consistent with the PDG average value $\alpha_s(M_Z)=0.1180\pm0.0009$~\cite{ParticleDataGroup:2024cfk} and other similar predictions within errors, e.g., a global fit to electro electroweak data gives $\alpha_s(M_Z)=0.1193\pm0.0028$~\cite{deBlas:2022hdk}, a global fit of jet rates at LEP and PETRA obtains $\alpha_s(M_Z)=0.1188\pm0.0013$ ~\cite{Verbytskyi:2019zhh}, and the Lattice QCD gives $\alpha_s(M_Z)=0.1184\pm0.0008$~\cite{FLAG:2021npn}.

\section{Summary}
\label{sec4}

In the present paper, we have proposed a new method to determine the magnitude of $\alpha_s$ at all scales with the help of BSR. The renormalization scheme-and-scale errors are always treated as an important systematic error of pQCD prediction. We adopt the $\rm{PMC}_\infty$ approach to deal with the pQCD corrections to $\alpha^{g_1}_s(Q)$, which reveals the intrinsic conformality of series and eliminates the conventional renormalization scheme-and-scale ambiguities. Thus the precision of pQCD prediction is greatly improved. Furthermore, to remove the previous ``self-consistence problem" emerged in $\overline{\rm MS}$-scheme series, we have adopted the $V$-scheme series to do our analysis. The resultant $\rm{PMC}_\infty$ series has only the \textit{the second kind of residual scale dependence}, which is negligible for the present case with up to four-loop QCD corrections. The LHFQCD model is used for the IR-behavior of $\alpha^{g_1}_s(Q)$. As a combination, we obtain a precise $\alpha^{g_1}_s(Q)$ at all scales, which agree well with the data with its $p$-value $\sim99\%$. We then determine the strong coupling constant at the scale $M_Z$, i.e. $\alpha_s(M_Z)=0.1191\pm{0.0012}\mp{0.0006}$, where the first error comes from $\Delta\kappa=\pm0.024$ GeV of LFHQCD model and the second error is from the unknown higher-order terms in pQCD series. It is consistent with the PDG average value within errors, thus being another example of showing the importance of PMC.

\acknowledgments{The authors extend their gratitude to Craig D. Roberts, Jose Rodr\'\i{}guez-Quintero and Daniele Binosi for sharing their DSE data in our discussions, and thank Xu-Dong Huang for the insightful discussions. This work was supported by the Natural Science Foundation of China under Grant No.12305091, No.12175025 and No.12347101, by the Natural Science Foundation of Sichuan Province under Grant No.2024NSFSC1367, and by the Research Fund for the Doctoral Program of the Southwest University of Science and Technology under Contract No.23zx7122 and No.24zx7117. J. M. Shen has been supported by YueLuShan Center for Industrial Innovation (2024YCII0118).

\appendix

\section*{Appendix: ransitions of the coefficients from $\overline{\rm MS}$-scheme to $V$-scheme}

The transition between the perturbative coefficients under $V$- and $\overline{\rm MS}$-scheme is utilizing the established relationships between the two coupling constants: $a^{\rm V}_s$ and $a^{\overline{\rm MS}}_s$. The two couplings under different schemes are satisfying the renormalization equation, i.e.,
\begin{eqnarray}
\beta^{\rm V}(a_s^{\rm V}) ={\partial a_s^{\rm V} \over \partial a_s^{\rm \overline{MS}}} \beta^{\rm \overline{MS}} (a_s^{\rm \overline{MS}}).
\end{eqnarray}
 The coefficients in the two $\beta$-functions are the same at the first two orders, i.e., $\beta_0={1\over4}(11-2n_f/3)$ and $\beta_1={1\over4^2}(102-38n_f/3)$ with quark flavor $n_f$. For present calculations, we need the four-loop QCD $\beta$-function under $V$-scheme~\cite{Kataev:2023sru,Kataev:2015yha} and $\overline{\rm MS}$-scheme~\cite{Czakon:2004bu,Chetyrkin:2004mf}, where the third and fourth terms are
\begin{eqnarray}
\beta^{\overline{\rm MS}}_2&=&\frac{1}{4^3}\left({2857\over2}-{5033\over18}n_f+{325\over54}n^2_f\right),\\
\beta^{\overline{\rm MS}}_3&=&\frac{1}{4^4}\left[{149753\over6}+3564\zeta_3-\left({1078361\over162}+{6508\over27}\zeta_3\right)n_f\right.\nonumber\\&&\left.
+\left({50065\over162}+{6472\over81}\zeta_3\right)n^2_f+{1093\over729}n^3_f\right],\\
\beta^V_2&=&\beta^{\rm \overline{MS}}_2-a_1\beta_1+(a_2-a^2_1)\beta_0,\\
\beta^V_3&=&\beta^{\rm \overline{MS}}_3-2a_1\beta^{\rm \overline{MS}}_2+a^2_1\beta_1+(2a_3-6a_1a_2+4a^3_1)\beta_0,\nonumber\\
\end{eqnarray}
where the known coefficients $a_i$~\cite{Kataev:2023sru,Kataev:2015yha} with quark flavor $n_f$ and Riemann zeta function $\zeta_3$ are
\begin{eqnarray}
a_1&=&{1\over4}({31\over3}-{10\over9}n_f),\\
a_2&=&{1\over4^2}\left[{4343\over18}+36\pi^2-{9\over4}\pi^4+66\zeta_3-({1229\over27}+{52\zeta_3\over3})n_f\right.\nonumber\\&&\left.
+{100\over81}n^2_f\right],\\
a_3&=&209.884-51.405n_f+2.906n^2_f-0.021n^3_f.
\end{eqnarray}
The transformation relation between $a_s^{\rm V}$ and $a_s^{\rm \overline{MS}}$ is known as~\cite{Chetyrkin:2004mf,Czakon:2004bu,Baikov:2016tgj,Kataev:2023sru}
\begin{eqnarray}
a^{\rm V}_{s}(\Lambda_{\rm V})&=&a^{\overline{\rm MS}}_s(\Lambda_{\rm V})\left[1+a_1a^{\overline{\rm MS}}_s(\Lambda_{\rm V})+a_2a^{\overline{\rm MS},2}_s(\Lambda_{\rm V})
\right.\nonumber\\&&\left.+a_3a^{\overline{\rm MS},3}_s(\Lambda_{\rm V})+\cdots\right],
\end{eqnarray}
where the strong coupling $a_s$ is a function of $Q$ and $\Lambda_{\rm QCD}$ in Eq.(\ref{alphas}). The symbol $\Lambda_{\rm V}$ is the $V$-scheme asymptotic scale which be defined as $\Lambda^{2}_{\rm V}=\Lambda^{2}_{\rm \overline{\rm MS}}\exp(a_1/\beta_0)$ with $\overline{\rm MS}$-scheme asymptotic scale $\Lambda_{\rm \overline{MS}}$.

The effective coupling $a^{g_1}_s(Q)$ in Eq.(\ref{ag1conv}) under $\overline{\rm MS}$-scheme includes the non-singlet and singlet contributions to the pQCD series up to $\rm N^3LO$~\cite{Gorishnii:1985xm,Larin:1991tj,Baikov:2010je,Larin:2013yba,Baikov:2015tea} and the decoupling mass corrections at $\mathcal{O}(a^{\overline{\rm MS},2}_s)$~\cite{Blumlein:2016xcy}. After complicating the transform into the $V$-scheme, the reduced pQCD coefficients $\hat{r}_{i,j}|_{Q=m_c}$ up to $\rm N^3LO$ in Eq.(\ref{ag1inf}) are
\begin{eqnarray}
{\hat r}^{\rm V}_{1,0} &=&\frac{3}{4}\gamma^{ns}_1, \\
{\hat r}^{\rm V}_{2,0} &=& \gamma^m_2+\frac{3}{2}\gamma^{ns}_1+\frac{3}{4}\gamma^{ns}_2-\frac{9}{16}(\gamma_1^{ns})^2, \\
{\hat r}^{\rm V}_{2,1} &=& -\frac{5}{4}\gamma^{ns}_1 +K^{ns}_1 +\frac{3}{4}\Pi^{ns}_1, \\
{\hat r}^{\rm V}_{3,0} &=&
4\gamma_2^m +3\gamma^{ns}_2+\frac{3}{4}\gamma_3^{ns}+3\times10^{-15}K_1^{ns}
\nonumber\\&&+\gamma^{ns}_1\left[11.796-\frac{9}{4}\gamma_1^{ns}-\frac{9}{8}\gamma_2^{ns}+ \frac{27}{64}(\gamma^{ns}_1)^2\right] , \nonumber\\
\\
{\hat r}^{\rm V}_{3,1} &=&-\frac{5}{3}\gamma_2^m-\frac{5}{4}\gamma_2^{ns}+2K_1^{ns} +\frac{K_2^{ns}}{2}+\frac{3}{2}\Pi^{ns}_1+\frac{3}{4}\Pi^{ns}_2
\nonumber\\&&-\gamma_1^{ns}\left(5.633-\frac{15}{16}\gamma_1^{ns}+\frac{3}{8}K_1^{ns}+\frac{9}{16}\Pi_1^{ns}\right), \nonumber\\
\\
{\hat r}^{\rm V}_{3,2} &=& \frac{25}{12}\gamma_1^{ns}-\frac{10}{3}K_1^{ns}-\frac{5}{2}\Pi_1^{ns}, \\
{\hat r}^{\rm V}_{4,0} &=&-10+35.456\gamma_2^m+26.592\gamma_2^{ns}+\frac{9}{2}\gamma_3^{ns}+\frac{3}{4}\gamma_4^{ns}\nonumber\\&&
-4\times10^{-14}K_1^{ns} -2\times10^{-14}\Pi_1^{ns}\nonumber\\&&
+\gamma_1^{ns}\left[6.832-\frac{27}{4}\gamma_2^{ns}-\frac{9}{8}\gamma_3^{ns}\right]-\frac{9}{16}(\gamma_2^{ns})^2\nonumber\\&&-(\gamma_1^{ns})^2\left[19.944-\frac{81}{64}\gamma_2^{ns}-\frac{81}{32}\gamma_1^{ns}+\frac{81}{256}(\gamma_1^{ns})^2\right], \nonumber\\  
\end{eqnarray}

\begin{eqnarray}
{\hat r}^{\rm V}_{4,1} &=& -12.237\gamma_2^m-9.178\gamma_2^{ns}-\frac{5}{4}\gamma_3^{ns} +2K_2^{ns} +\frac{K_3^{ns}}{3}  \nonumber\\&&
+3\Pi_2^{ns}+\frac{3}{4}\Pi_3^{ns}-\gamma_1^{ns}\left(16.163-\frac{15}{8}\gamma_2^{ns}+\frac{3}{2}K_1^{ns}\right.\nonumber\\&&\left.+\frac{K_2^{ns}}{4}+\frac{9}{4}\Pi_1^{ns}
+\frac{3}{4}\Pi_2^{ns}\right)
+(\gamma_1^{ns})^2\left[6.883+\frac{3}{16}K_1^{ns}
\right.\nonumber\\&&\left.
+\frac{27}{64}\Pi_1^{ns}-\frac{45}{64}\gamma_1^{ns}\right]+K_1^{ns}\left(15.728-\frac{\gamma_2^{ns}}{4}\right)
\nonumber\\&&
+\Pi_1^{ns}\left(11.796-\frac{3}{8}\gamma_2^{ns}\right),\\
{\hat r}^{\rm V}_{4,2} &=& \frac{25}{9}\gamma_2^m+\frac{25}{12}\gamma_2^{ns}-15.022 K_1^{ns}-\frac{5}{3}K_2^{ns}-\frac{5}{2}\Pi_2^{ns} \nonumber\\&&
+\gamma_1^{ns}\left[13.939-\frac{25}{16}\gamma_1^{ns}+\frac{5}{4}K_1^{ns}+\frac{15}{8}\Pi_1^{ns}\right]\nonumber\\&&
-\Pi_1^{ns}\left[11.266+\frac{K_1^{ns}}{4}+\frac{3}{16}\Pi_1^{ns}\right],\\
{\hat r}^{\rm V}_{4,3} &=& -\frac{125}{36}\gamma_1^{ns}+\frac{25}{3}K_1^{ns}+\frac{25}{4}\Pi_1^{ns}.
\end{eqnarray}
where $\gamma_i^{ns}$-, $K_i^{ns}$ and $\Pi_i^{ns}$-functions defined in Refs.\cite{Baikov:2010je, Baikov:2015tea}. It has been emphasized that only the RGE-involved $n_f$-terms should be adopted for fixing the magnitude of $\alpha_s$, and for the present case, the $n_f$-terms pertained to anomalous dimension should be teated as conformal terms~\cite{Shen:2017pdu, Yan:2023hra}~\footnote{There are some wrong comments on PMC which are based on using all $n_f$-terms. Practically, when we cannot do strict distinguishing of whether the $n_f$-terms are RGE-involved or uninvolved, such a simple treatment could be treated as an effective PMC treatment, which may also works well.}. And the decoupling mass corrections~\cite{Blumlein:2016xcy} are
\begin{eqnarray}
\gamma^m_2&=&C^m_{\rm pBJ}(\xi_c)+C^m_{\rm pBJ}(\xi_b),\\
C^m_{\rm pBJ}&=&\frac{2}{3}\left[\frac{\xi_x^2+2735\xi+11724}{5040\xi_x}-\frac{\sqrt{\xi_x+4}}{5040\xi_x^{3/2}}\left(3\xi_x^3
\right.\right.\nonumber\\&&\left.\left.
+106\xi_x^2+1054\xi_x+4812\right)\log(\frac{\sqrt{\xi_x+4}+\sqrt{\xi_x}}{\sqrt{\xi_x+4}-\sqrt{\xi_x}})
\right.\nonumber\\&&\left.
-\frac{5}{12\xi_x^2}
\log^2(\frac{\sqrt{\xi_x+4}+\sqrt{\xi_x}}{\sqrt{\xi_x+4}-\sqrt{\xi_x}})
\right.\nonumber\\&&\left.
+\log(\xi_x)\frac{3\xi_x^2+112\xi_x+1260}{5040}\right],\nonumber\\
\end{eqnarray}
where $\xi_c=Q^2/m^2_c$ and $\xi_b=Q^2/m^2_b$.


\begin{thebibliography}{99}

\bibitem{Gross:1973id}
D.~J.~Gross and F.~Wilczek,
Phys. Rev. Lett. \textbf{30}, 1343 (1973).

\bibitem{Politzer:1973fx}
H.~D.~Politzer,
Phys. Rev. Lett. \textbf{30}, 1346 (1973).

\bibitem{tHooft:1973mfk}
G.~'t Hooft,
Nucl. Phys. B \textbf{61}, 455 (1973).

\bibitem{Weinberg:1973xwm}
S.~Weinberg,
Phys. Rev. D \textbf{8}, 3497 (1973).

\bibitem{Bjorken:1966jh}
  J.~D.~Bjorken,
  Phys.\ Rev.\  {\bf 148}, 1467 (1966).

\bibitem{Bjorken:1969mm}
  J.~D.~Bjorken,
  Phys.\ Rev.\ D {\bf 1}, 1376 (1970).

\bibitem{Ackerstaff:1997ws}
 K.~Ackerstaff {\it et al.} [HERMES Collaboration],
Phys.\ Lett.\ B {\bf 404}, 383 (1997);
 K.~Ackerstaff {\it et al.} [HERMES Collaboration],
Phys.\ Lett.\ B {\bf 444}, 531 (1998);
 A.~Airapetian {\it et al.} [HERMES Collaboration],
 Phys.\ Lett.\ B {\bf 442}, 484 (1998);
  A.~Airapetian {\it et al.} [HERMES Collaboration],
  Phys.\ Rev.\ Lett.\  {\bf 90}, 092002 (2003);
  A.~Airapetian {\it et al.} [HERMES Collaboration],
  Phys.\ Rev.\ D {\bf 75}, 012007 (2007).

\bibitem{Alexakhin:2006oza}
 V.~Y.~Alexakhin {\it et al.} [COMPASS Collaboration],
 Phys.\ Lett.\ B {\bf 647}, 8 (2007);
M.~G.~Alekseev {\it et al.} [COMPASS Collaboration],
 Phys.\ Lett.\ B {\bf 690}, 466 (2010);
  C.~Adolph {\it et al.} [COMPASS Collaboration],
  Phys.\ Lett.\ B {\bf 753}, 18 (2016).

\bibitem{Anthony:1993uf}
  P.~L.~Anthony {\it et al.} [E142 Collaboration],
  Phys.\ Rev.\ Lett.\  {\bf 71}, 959 (1993);
  P.~L.~Anthony {\it et al.} [E142 Collaboration],
  Phys.\ Rev.\ D {\bf 54}, 6620 (1996);
  K.~Abe {\it et al.} [E143 Collaboration],
  Phys.\ Rev.\ Lett.\  {\bf 74}, 346 (1995);
  K.~Abe {\it et al.} [E143 Collaboration],
  Phys.\ Rev.\ Lett.\  {\bf 75}, 25 (1995);
  K.~Abe {\it et al.} [E143 Collaboration],
  Phys.\ Rev.\ Lett.\  {\bf 76}, 587 (1996);
  K.~Abe {\it et al.} [E143 Collaboration],
  Phys.\ Lett.\ B {\bf 364}, 61 (1995);
  K.~Abe {\it et al.} [E143 Collaboration],
  Phys.\ Rev.\ D {\bf 58}, 112003 (1998).

\bibitem{Deur:2004ti}
  A.~Deur {\it et al.},
  Phys.\ Rev.\ Lett.\  {\bf 93}, 212001 (2004).

\bibitem{Deur:2005cf}
  A.~Deur, V.~Burkert, J.~P.~Chen and W.~Korsch,
  Phys. Lett. B \textbf{650}, 244 (2007).

\bibitem{Deur:2008ej}
  A.~Deur {\it et al.},
  Phys.\ Rev.\ D {\bf 78}, 032001 (2008).

\bibitem{Deur:2014vea}
  A.~Deur {\it et al.},
  Phys.\ Rev.\ D {\bf 90}, 012009 (2014).

\bibitem{Deur:2022msf}
A.~Deur, V.~Burkert, J.~P.~Chen and W.~Korsch,
Particles \textbf{5}, 171 (2022).

\bibitem{ParticleDataGroup:2024cfk}
S.~Navas \textit{et al.} [Particle Data Group],
Phys. Rev. D \textbf{110}, 030001 (2024).

\bibitem{Gorishnii:1985xm}
S.~G.~Gorishnii and S.~A.~Larin,
Phys. Lett. B \textbf{172}, 109 (1986).

\bibitem{Larin:1991tj}
S.~A.~Larin and J.~A.~M.~Vermaseren,
Phys. Lett. B \textbf{259}, 345 (1991).

\bibitem{Baikov:2010je}
  P.~A.~Baikov, K.~G.~Chetyrkin and J.~H.~Kuhn,
  Phys.\ Rev.\ Lett.\  {\bf 104}, 132004 (2010).

\bibitem{Larin:2013yba}
S.~A.~Larin,
Phys. Lett. B \textbf{723}, 348 (2013).

\bibitem{Baikov:2015tea}
P.~A.~Baikov, K.~G.~Chetyrkin and J.~H.~K\"uhn,
Nucl. Part. Phys. Proc. \textbf{261-262}, 3 (2015).

\bibitem{Blumlein:2016xcy}
J.~Bl\"umlein, G.~Falcioni and A.~De Freitas,
Nucl. Phys. B \textbf{910}, 568 (2016).

\bibitem{Ayala:2018ulm}
C.~Ayala, G.~Cveti\v{c}, A.~V.~Kotikov and B.~G.~Shaikhatdenov,
Eur. Phys. J. C \textbf{78}, 1002 (2018).

\bibitem{Yu:2021ofs}
  Q.~Yu, X.~G.~Wu, H.~Zhou and X.~D.~Huang,
  Eur. Phys. J. C \textbf{81}, 690 (2021).

\bibitem{Ayala:2023cxm}
C.~Ayala, C.~Castro-Arriaza and G.~Cvetic,
Nucl. Phys. B \textbf{1007}, 116668 (2024)

\bibitem{Brodsky:2014yha}
  S.~J.~Brodsky, G.~F.~de Teramond, H.~G.~Dosch and J.~Erlich,
  Phys.\ Rept.\  {\bf 584}, 1 (2015).

\bibitem{Binosi:2016nme}
D.~Binosi, C.~Mezrag, J.~Papavassiliou, C.~D.~Roberts and J.~Rodriguez-Quintero,
Phys. Rev. D \textbf{96}, 054026 (2017).

\bibitem{Cui:2019dwv}
Z.~F.~Cui, J.~L.~Zhang, D.~Binosi, F.~de Soto, C.~Mezrag, J.~Papavassiliou, C.~D.~Roberts, J.~Rodr\'\i{}guez-Quintero, J.~Segovia and S.~Zafeiropoulos,
Chin. Phys. C \textbf{44}, 083102 (2020).

\bibitem{Deur:2023dzc}
A.~Deur, S.~J.~Brodsky and C.~D.~Roberts,
Prog. Part. Nucl. Phys. \textbf{134}, 104081 (2024).

\bibitem{Brodsky:2024zev}
S.~J.~Brodsky, A.~Deur and C.~D.~Roberts,
Scientific American, 32 (2024).

\bibitem{Petermann:1953wpa}
  A.~Petermann,
  Helv.\ Phys.\ Acta {\bf 26}, 499 (1953).

\bibitem{GellMann:1954fq}
  M.~Gell-Mann and F.~E.~Low,
  Phys.\ Rev.\  {\bf 95}, 1300 (1954).

\bibitem{Callan:1970yg}
  C.~G.~Callan, Jr.,
  Phys.\ Rev.\ D {\bf 2}, 1541 (1970).

\bibitem{Symanzik:1970rt}
  K.~Symanzik,
  Commun.\ Math.\ Phys.\  {\bf 18}, 227 (1970).

\bibitem{Peterman:1978tb}
  A.~Peterman,
  Phys.\ Rept.\  {\bf 53}, 157 (1979).

\bibitem{Brodsky:1982gc}
 S.~J.~Brodsky, G.~P.~Lepage and P.~B.~Mackenzie,
 Phys. Rev. D \textbf{28}, 228 (1983).

\bibitem{Yu:2021yvw}
Q.~Yu, H.~Zhou, X.~D.~Huang, J.~M.~Shen and X.~G.~Wu,
Chin. Phys. Lett. \textbf{39}, 071201 (2022).

\bibitem{Brodsky:2011ta}
  S.~J.~Brodsky and X.~G.~Wu,
  Phys.\ Rev.\ D {\bf 85}, 034038 (2012).

\bibitem{Brodsky:2012rj}
  S.~J.~Brodsky and X.~G.~Wu,
  Phys.\ Rev.\ Lett.\  {\bf 109}, 042002 (2012).

\bibitem{Brodsky:2011ig}
  S.~J.~Brodsky and L.~Di Giustino,
  Phys.\ Rev.\ D {\bf 86}, 085026 (2012).

\bibitem{Mojaza:2012mf}
  M.~Mojaza, S.~J.~Brodsky and X.~G.~Wu,
  Phys.\ Rev.\ Lett.\  {\bf 110}, 192001 (2013).

\bibitem{Brodsky:2013vpa}
  S.~J.~Brodsky, M.~Mojaza and X.~G.~Wu,
  Phys.\ Rev.\ D {\bf 89}, 014027 (2014).

\bibitem{Wu:2014iba}
  X.~G.~Wu {\it et al.},
  Rept. Prog. Phys. \textbf{78}, 126201 (2015).


\bibitem{Wu:2018cmb}
  X.~G.~Wu, J.~M.~Shen, B.~L.~Du and S.~J.~Brodsky,
  Phys.\ Rev.\ D {\bf 97}, 094030 (2018).

\bibitem{Wu:2019mky}
  X.~G.~Wu, J.~M.~Shen, B.~L.~Du, X.~D.~Huang, S.~Q.~Wang and S.~J.~Brodsky,
  Prog.\ Part.\ Nucl.\ Phys.\  {\bf 108}, 103706 (2019).

\bibitem{Zheng:2013uja}
X.~C.~Zheng, X.~G.~Wu, S.~Q.~Wang, J.~M.~Shen and Q.~L.~Zhang,
JHEP \textbf{10}, 117 (2013).

\bibitem{Brodsky:2010ur}
  S.~J.~Brodsky, G.~F.~de Teramond and A.~Deur,
  Phys.\ Rev.\ D {\bf 81}, 096010 (2010).

\bibitem{Deur:2014qfa}
  A.~Deur, S.~J.~Brodsky and G.~F.~de Teramond,
  Phys.\ Lett.\ B {\bf 750}, 528 (2015).

\bibitem{Deur:2016cxb}
  A.~Deur, S.~J.~Brodsky and G.~F.~de Teramond,
  Phys.\ Lett.\ B {\bf 757}, 275 (2016).

 \bibitem{Deur:2017cvd}
  A.~Deur, J.~M.~Shen, X.~G.~Wu, S.~J.~Brodsky and G.~F.~de Teramond,
  Phys.\ Lett.\ B {\bf 773}, 98 (2017).

\bibitem{Shen:2017pdu}
  J.~M.~Shen, X.~G.~Wu, B.~L.~Du and S.~J.~Brodsky,
  Phys.\ Rev.\ D {\bf 95}, 094006 (2017).

  \bibitem{DiGiustino:2020fbk}
L.~Di Giustino, S.~J.~Brodsky, S.~Q.~Wang and X.~G.~Wu,
Phys. Rev. D \textbf{102}, 014015 (2020).

\bibitem{Peter:1996ig}
  M.~Peter,
  Phys.\ Rev.\ Lett.\  {\bf 78}, 602 (1997).

\bibitem{Schroder:1998vy}
  Y.~Schroder,
  Phys.\ Lett.\ B {\bf 447}, 321 (1999).

\bibitem{Appelquist:1977tw}
  T.~Appelquist, M.~Dine and I.~J.~Muzinich,
  Phys.\ Lett.\ B {\bf 69}, 231 (1977).

\bibitem{Fischler:1977yf}
  W.~Fischler,
  Nucl.\ Phys.\ B {\bf 129}, 157 (1977).

\bibitem{Billoire:1979ih}
A.~Billoire,
Phys. Lett. B \textbf{92}, 343 (1980).

\bibitem{Huang:2021hzr}
X.~D.~Huang, J.~Yan, H.~H.~Ma, L.~Di Giustino, J.~M.~Shen, X.~G.~Wu and S.~J.~Brodsky,
Nucl. Phys. B \textbf{989}, 116150 (2023).

\bibitem{Kataev:2015yha}
A.~L.~Kataev and V.~S.~Molokoedov,
Phys. Rev. D \textbf{92}, 054008 (2015).

\bibitem{Kataev:2023sru}
A.~L.~Kataev and V.~S.~Molokoedov,
Theor. Math. Phys. \textbf{217}, 1459 (2023).

\bibitem{Chetyrkin:2004mf}
  K.~G.~Chetyrkin,
  Nucl.\ Phys.\ B {\bf 710}, 499 (2005).

\bibitem{Czakon:2004bu}
  M.~Czakon,
  Nucl.\ Phys.\ B {\bf 710}, 485 (2005).

\bibitem{Baikov:2016tgj}
  P.~A.~Baikov, K.~G.~Chetyrkin and J.~H.~Khn,
  Phys.\ Rev.\ Lett.\  {\bf 118}, 082002 (2017).

\bibitem{Yan:2023hra}
J.~Yan, S.~J.~Brodsky, L.~Di Giustino, P.~G.~Ratcliffe, S.~Q.~Wang and X.~G.~Wu,
[arXiv:2311.17360 [hep-ph]].

\bibitem{Brodsky:1994eh}
S.~J.~Brodsky and H.~J.~Lu,
Phys. Rev. D \textbf{51}, 3652 (1995).


\bibitem{Basdevant:1972fe}
  J.~L.~Basdevant,
  Fortsch.\ Phys.\  {\bf 20}, 283 (1972).

\bibitem{Chetyrkin:1997sg}
K.~G.~Chetyrkin, B.~A.~Kniehl and M.~Steinhauser,
Phys. Rev. Lett. \textbf{79}, 2184 (1997).

\bibitem{Du:2018dma}
  B.~L.~Du, X.~G.~Wu, J.~M.~Shen and S.~J.~Brodsky,
  Eur.\ Phys.\ J.\ C {\bf 79}, 182 (2019).

\bibitem{Prosperi:2006hx}
  G.~M.~Prosperi, M.~Raciti and C.~Simolo,
  Prog.\ Part.\ Nucl.\ Phys.\  {\bf 58}, 387 (2007).

\bibitem{Deur:2016tte}
  A.~Deur, S.~J.~Brodsky and G.~F.~de Teramond,
  Prog.\ Part.\ Nucl.\ Phys.\  {\bf 90}, 1 (2016).

\bibitem{Brodsky:2016yod}
S.~J.~Brodsky, G.~F.~de T\'eramond, H.~G.~Dosch and C.~Lorc\'e,
Phys. Lett. B \textbf{759} (2016), 171-177.

\bibitem{Ackerstaff:1998yj}
  K.~Ackerstaff {\it et al.} [OPAL Collaboration],
  Eur.\ Phys.\ J.\ C {\bf 7}, 571 (1999).

\bibitem{Brodsky:2002nb}
  S.~J.~Brodsky, S.~Menke, C.~Merino and J.~Rathsman,
  Phys.\ Rev.\ D {\bf 67}, 055008 (2003).

\bibitem{Gross:1969jf}
  D.~J.~Gross and C.~H.~Llewellyn Smith,
  Nucl.\ Phys.\ B {\bf 14}, 337 (1969).

\bibitem{Kim:1998kia}
  J.~H.~Kim {\it et al.},
  Phys.\ Rev.\ Lett.\  {\bf 81}, 3595 (1998).

\bibitem{Abe:1997cx}
  K.~Abe {\it et al.} [E154 Collaboration],
  Phys.\ Rev.\ Lett.\  {\bf 79}, 26 (1997);
  K.~Abe {\it et al.} [E154 Collaboration],
  Phys.\ Lett.\ B {\bf 404}, 377 (1997);
  K.~Abe {\it et al.} [E154 Collaboration],
  Phys.\ Lett.\ B {\bf 405}, 180 (1997).

\bibitem{Anthony:1999py}
 P.~L.~Anthony {\it et al.} [E155 Collaboration],
  Phys.\ Lett.\ B {\bf 458}, 529 (1999);
  P.~L.~Anthony {\it et al.} [E155 Collaboration],
  Phys.\ Lett.\ B {\bf 463}, 339 (1999);
  P.~L.~Anthony {\it et al.} [E155 Collaboration],
  Phys.\ Lett.\ B {\bf 493}, 19 (2000);
  P.~L.~Anthony {\it et al.} [E155 Collaboration],
  Phys.\ Lett.\ B {\bf 553}, 18 (2003).

\bibitem{Adeva:1998vv}
  B.~Adeva {\it et al.} [Spin Muon Collaboration],
  Phys.\ Rev.\ D {\bf 58}, 112001 (1998).

 \bibitem{Zafeiropoulos:2019flq}
S.~Zafeiropoulos, P.~Boucaud, F.~De Soto, J.~Rodr\'\i{}guez-Quintero and J.~Segovia,
Phys. Rev. Lett. \textbf{122}, 162002 (2019).

\bibitem{Shen:2023qgz}
J.~M.~Shen, B.~H.~Qin, J.~Yan, S.~Q.~Wang and X.~G.~Wu,
JHEP \textbf{07}, 109 (2023).

\bibitem{Proceedings:2019pra}
D.~d'Enterria, S.~Kluth, S.~Alekhin, P.~A.~Baikov, A.~Banfi, F.~Barreiro, A.~Bazavov, S.~Bethke, J.~Bl\"umlein and D.~Boito, \textit{et al.}
[arXiv:1907.01435 [hep-ph]].

\bibitem{dEnterria:2022hzv}
D.~d'Enterria, S.~Kluth, G.~Zanderighi, C.~Ayala, M.~A.~Benitez-Rathgeb, J.~Bluemlein, D.~Boito, N.~Brambilla, D.~Britzger and S.~Camarda, \textit{et al.}
[arXiv:2203.08271 [hep-ph]].

\bibitem{deBlas:2022hdk}
J.~de Blas, M.~Pierini, L.~Reina and L.~Silvestrini,
Phys. Rev. Lett. \textbf{129}, 271801 (2022).

\bibitem{Verbytskyi:2019zhh}
A.~Verbytskyi, A.~Banfi, A.~Kardos, P.~F.~Monni, S.~Kluth, G.~Somogyi, Z.~Sz\H{o}r, Z.~Tr\'ocs\'anyi, Z.~Tulip\'ant and G.~Zanderighi,
JHEP \textbf{08}, 129 (2019).


\bibitem{FLAG:2021npn}
Y.~Aoki \textit{et al.} [FLAG],
Eur. Phys. J. C \textbf{82}, 869 (2022).


\bibitem{CMS:2024mlf}
A.~Hayrapetyan \textit{et al.} [CMS],
Phys. Rev. Lett. \textbf{133}, 071903 (2024).



\end{thebibliography}
\end{document}